\documentclass[apj]{emulateapj-rtx4}

\usepackage{calc}
\shorttitle{SN~2010as and Flat-Velocity SN~IIb}
\shortauthors{Folatelli et al.}

\begin{document}

\title{Supernova 2010as: the Lowest-Velocity Member of a Family of Flat-Velocity Type IIb Supernovae\altaffilmark{1}}

\author{%
  Gast\'on Folatelli\altaffilmark{2},
  Melina C. Bersten\altaffilmark{2},
  Hanindyo Kuncarayakti\altaffilmark{3,4},
  Felipe Olivares Estay\altaffilmark{3,5},
  Joseph P. Anderson\altaffilmark{6},
  Simon Holmbo\altaffilmark{7},
  Keiichi Maeda\altaffilmark{8},
  Nidia Morrell\altaffilmark{9},
  Ken'ichi Nomoto\altaffilmark{2,10},
  Giuliano Pignata\altaffilmark{5,3},
  Maximilian Stritzinger\altaffilmark{7},
  Carlos Contreras\altaffilmark{9},
  Francisco F\"orster\altaffilmark{11,3},
  Mario Hamuy\altaffilmark{4,3},
  Mark M. Phillips\altaffilmark{9},
  Jos\'e Luis Prieto\altaffilmark{12,13},
  Stefano Valenti\altaffilmark{14},
  Paulo Afonso\altaffilmark{15},
  Konrad Altenm\"uller\altaffilmark{15,16}, 
  Jonny Elliott\altaffilmark{15},
  Jochen Greiner\altaffilmark{15},
  Adria Updike\altaffilmark{17},
  Joshua B. Haislip\altaffilmark{18},
  Aaron P. LaCluyze\altaffilmark{18},
  Justin P. Moore\altaffilmark{18},
  and
  Daniel E. Reichart\altaffilmark{18}
}
\altaffiltext{1}{This paper includes data gathered with the following
  facilities in Chile: the 6.5~m Magellan Telescopes located at Las
  Campanas Observatory, the Gemini Observatory, Cerro Pach\'on (Gemini Program
  GS-2008B$-$Q$-$56), and the European Organisation for Astronomical
  Research in the Southern Hemisphere (ESO Programmes 076.A-0156,
  078.D-0048, 080.A-0516, and 082.A-0526). We have also used data from
the ESO Science Archive Facility under request number
gfolatelli74580, and from the NASA/ESA Hubble Space Telescope,
obtained from the Hubble Legacy Archive, which is a collaboration
between the Space Telescope Science Institute (STScI/NASA), the Space
Telescope European Coordinating Facility (ST-ECF/ESA) and the Canadian
Astronomy Data Centre (CADC/NRC/CSA).} 
\altaffiltext{2}{Kavli Institute for the Physics and Mathematics of
  the Universe (WPI), The University of Tokyo, Kashiwa, Chiba
  277-8583, Japan}
\altaffiltext{3}{Millennium Institute of Astrophysics (MAS), Casilla
  36-D, Santiago, Chile}  
\altaffiltext{4}{Departamento de Astronom\'ia, Universidad de Chile,
  Casilla 36-D, Santiago, Chile}  
\altaffiltext{5}{Departamento de Ciencias Fisicas, Universidad Andres
  Bello, Avda.\ Republica 252, Santiago, Chile} 
\altaffiltext{6}{European Southern Observatory, Alonso de Cordova
  3107, Vitacura, Santiago, Chile} 
\altaffiltext{7}{Department of Physics and Astronomy, Aarhus
  University, Ny Munkegade 120, DK-8000 Aarhus C, Denmark}
\altaffiltext{8}{Department of Astronomy, Kyoto University,
  Kitashirakawa-Oiwake-cho, Sakyo-ku, Kyoto 606-8502, Japan} 
\altaffiltext{9}{Las Campanas Observatory, Carnegie Observatories,
  Casilla 601, La Serena, Chile} 
\altaffiltext{10}{Hamamatsu Professor}
\altaffiltext{11}{Center for Mathematical Modelling, Universidad de
  Chile, Avenida Blanco Encalada 2120 Piso 7, Santiago, Chile} 
\altaffiltext{12}{Department of Astrophysical Sciences, Princeton
  University, 4 Ivy Lane, Peyton Hall, Princeton, NJ 08544, USA} 
\altaffiltext{13}{N\'ucleo de Astronom\'ia de la Facultad de
  Ingenier\'ia, Universidad Diego Portales, Av. Ej\'ercito 441,
  Santiago, Chile} 
\altaffiltext{14}{Las Cumbres Observatory Global Telescope Network,
  6740 Cortona Dr., Suite 102, Goleta, CA 93117, USA} 
\altaffiltext{15}{Max-Planck-Institut f\"ur extraterrestrische Physik,
 Giessenbachstra\ss{}e 1, 85740 Garching, Germany}
\altaffiltext{16}{Technische Universit\"at M\"unchen, Physik
 Department, James-Franck-Strasse, 85748 Garching, Germany} 
\altaffiltext{17}{Department of Astronomy, University of Maryland,
  College Park, MD 20742, USA} 
\altaffiltext{18}{Department of Physics and Astronomy, University of
  North Carolina at Chapel Hill, Chapel Hill, NC 27599-3255} 

\email{gaston.folatelli@ipmu.jp}

\setcounter{footnote}{19}

\begin{abstract}
\noindent We present extensive optical and near-infrared photometric and
spectroscopic observations of the stripped-envelope (SE)
supernova SN~2010as. Spectroscopic peculiarities, such as initially weak
helium features and low expansion velocities with a nearly flat evolution,
place this object in the small family of events previously identified
as transitional Type~Ib/c supernovae (SNe). There is ubiquitous evidence of
hydrogen, albeit weak, in this family of SNe, indicating that they are
in fact a peculiar kind of Type~IIb SNe that we name ``flat-velocity
Type~IIb''. The flat velocity evolution---which occurs
at different levels between 6000 and 8000 km 
s$^{-1}$ for different SNe---suggests the 
presence of a dense shell in the ejecta. Despite the spectroscopic
similarities, these objects show surprisingly diverse luminosities. 
We discuss the possible physical or geometrical unification picture
for such diversity. Using archival {\em HST} images we associate
SN~2010as with a massive cluster and derive a progenitor age of
$\approx$6 Myr, assuming a single star-formation burst, which is
compatible with a Wolf-Rayet progenitor. Our hydrodynamical modelling,
on the contrary, indicates the pre-explosion mass was relatively
low, of $\approx$4 $M_\odot$. The seeming contradiction between an
young age and low pre-SN mass may be solved by a massive interacting
binary progenitor. 
\end{abstract}


\section{INTRODUCTION}
\label{sec:intro}

\noindent Stripped-envelope supernovae (SE SNe)---i.e., those with
hydrogen-poor spectra---are observationally very diverse. This
somewhat loosely defined group comprises H-poor Type~IIb SNe, H-deficient
and He-rich Type~Ib SNe, and both H- and He-deficient Type~Ic
SNe. While it is established that most of these events arise from the
core collapse of massive stars (with initial masses above 8
$M_\odot$), we lack a clear knowledge concerning the connection
between different  
subtypes and stellar progenitors or explosion processes. In particular,
there is the long-standing question of how stars get rid of their
outer envelopes before exploding. Is it through strong winds, by
eruptive episodes like in luminous blue variable stars, by mass 
transfer in close binary systems, or are there other mechanisms at
play \citep[see, e.g.,][]{langer12}? Is there a distinct separation
between Type~Ib and 
Type~IIb SNe, or is it a continuum regulated by the amount of hydrogen
\citep{nomoto95,heger03,georgy09,yoon10}? What is the reason for
Type~Ic SNe not to 
show helium in the spectra \citep{hachinger12,dessart12a}? Theory and
observation can help each other 
to solve these issues and thereby provide a more physically
motivated subdivision of SE SN types. 

Some pieces of the puzzle can be sought for in the increasing amount
of events intensively observed soon after explosion. For
instance, this has led to the discovery of SNe with observational
properties in between the subtype definitions. Of particular interest
among these is a small group of SNe with spectroscopic characteristics
that drift from Type~Ic to Type~Ib as helium lines become more prevalent
with phase. This sort of behavior was first identified for SN~1999ex 
\citep{hamuy02}, and a subclass of intermediate or
transitional Type~Ib/c was introduced. The term transitional is used
here to distinguish this classification from a loose ``Type~Ibc'' 
class that is sometimes used to indicate that the SN could either be
Type~Ib or Type~Ic. Similar cases were subsequently found, such as SN~2005bf
\citep{anupama05,tominaga05,folatelli06}, SN~2007Y
\citep{stritzinger09}, and SNe~2006lc and 2007kj 
\citep{leloudas11}. Besides the weakness of helium lines, a 
defining characteristic of this group is the persistence of low
helium-line expansion velocities, as shown for SN~1999ex and SN~2007Y
\citep{taubenberger11}, and for SN~2005bf \citep{tominaga05}. Hints of
hydrogen in the spectra have also been pointed out for these objects. 
In fact, the spectral typing of these objects in the literature is quite
ambiguous. Moreover, their photometric properties are very diverse,
including the low luminosity of SN~2007Y and the very luminous and
peculiar re-brightening of SN~2005bf. The question is what physical
conditions or observational biases may cause the spectroscopic
similarities to be accompanied by such heterogeneity in photometric
properties. 

SN~2010as joins this small family of transitional
Type~Ib/c SNe. The Millennium Center for Supernova Studies
\citep[MCSS;][]{hamuy12}\footnote{MCSS has been enlarged to form the
  Millennium institute of AStrophysics (MAS).} and 
collaborators collected a large amount of multi-wavelength data
following the SN evolution. This paper presents an analysis of the
observed spectroscopic and photometric properties of SN~2010as. We
ascertain the identification of spectral features based on SYNOW
synthetic spectra. Hydrodynamical modelling of the bolometric light
curve is used for deriving information about the progenitor star and explosion
characteristics. We also study the environmental properties near the
SN site from deep spectroscopy. Available pre-explosion Hubble Space
Telescope ({\em HST\,}) imaging provides a rare opportunity of
studying the underlying stellar population and consequently of
inferring an age of the progenitor. With the excellent data set collected
for SN~2010as and the comparison with spectroscopically similar SNe we
review the spectral classification of this group of objects.

The observations are described in Section~\ref{sec:obs}
and analyzed in Section~\ref{sec:ana}. Section~\ref{sec:prog}
focuses on the possible properties of the progenitor of SN~2010as, from
population analysis and hydrodynamical modelling. In
Section~\ref{sec:family} we identify a class of SE SNe with peculiar
velocities that includes SN~2010as, and discuss the possible origin
of the velocity behavior. We finalize with our conclusions in
Section~\ref{sec:concl}. 

\section{OBSERVATIONS}
\label{sec:obs}

\begin{figure}[htpb] % Figure 1
\epsscale{1.0}
\plotone{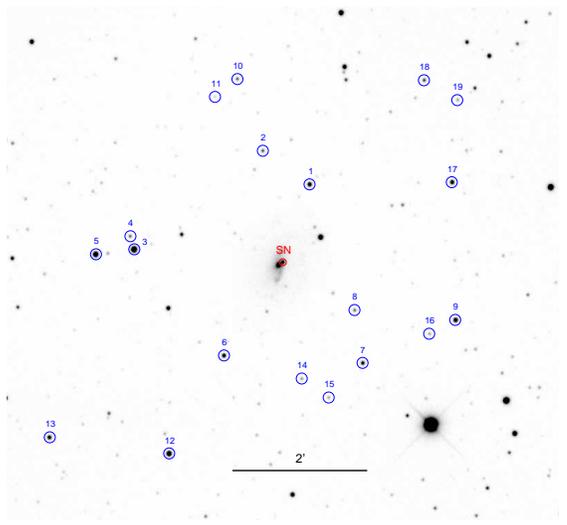}
\caption{CHASE $V$-band image of the field of NGC~6000 and
  SN~2010as. A red circle shows the location of the SN. Numbered blue 
  circles indicate the stars used for photometric calibration (see
  Tables~\ref{tab:lseq}, \ref{tab:sseq} and \ref{tab:nseq}). The scale
  of the image is indicated by the horizontal bar. North is up and
  East is to the left. A close-up of the host galaxy and SN~2010as as
  observed by VLT with NaCo in the $K$ band can be seen in
  Figure~\ref{fig:presn}. 
  \label{fig:fc}}
\end{figure}

SN~2010as was discovered by the CHilean Automatic Supernova sEarch
\citep[CHASE;][]{pignata09} on 2010 March $19.22$ UT \citep{maza10} using the
PROMPT~1 telescope installed on the Cerro Tololo Interamerican
Observatory (CTIO), Chile. The object was located very near the
nucleus of the starburst galaxy NGC~6000, which had also hosted the Type II
SN~2007ch. With coordinates $\alpha = 15^{\mathrm{h}}49^{\mathrm{m}}49{\fs}23$, 
$\delta=-29^\circ23'09{\farcs}7$, the SN was only $3{\farcs}1$ West
and $2{\farcs}0$ North of the center of NGC~6000. Figure~\ref{fig:fc}
shows a chart of the SN field. 

Prior to the discovery, the field of NGC~6000 was observed by CHASE on
2010 March $12.33$ UT with no detection of any source to a limiting
unfiltered magnitude of $\approx 17.5$ mag. We thus can constrain the
explosion date between March $12.3$ and March $19.2$---i.e., between
JD\,$=2455267.8$ and $2455274.7$.

Spectroscopy obtained soon after the discovery indicated that the new
SN was similar to objects such as SN~2007Y at a very early phase
\citep{stritzinger10}, and it was thus classified as a transitional
Type Ib/c SN. Given the proximity and young age of the SN, intensive
follow-up was started from several facilities. The observations are
described in the rest of the current section.

\subsection{Photometry}
\label{sec:phot}

\noindent Multi-band optical imaging was obtained using the PROMPT
$0.41$~m telescopes at CTIO following the SN evolution for over
100 days. Both Johnson-Kron-Cousins $BVRI$ bands and Sloan $g'r'i'z'$ bands
were used. We observed in $BVRI$ with the PROMPT~1 telescope, in
$B$ and $g'$ with PROMPT~3, and in $VRI$ and $r'i'z'$ with
PROMPT~5. The images covered a field of view of $\approx$ 
$10' \times 10'$, with a scale of $0{\farcs}6$ per pixel. Because of
lack of accurate guiding of the telescopes, the integration time was
set to 40 seconds. Integrations were repeated with each filter in
order to obtain high enough signal-to-noise ratios. Each integration
was corrected for dark current and flat-fielded before being
registered and stacked into a single image per filter. 

Because the SN appeared on a bright and complex background, we
performed image subtraction in order to obtain accurate
photometry. For this purpose we obtained deep reference images in all
bands with the PROMPT telescopes between 2011 March and 2012 March,
i.e. at least 350 days after the discovery of the SN, and thus when
its flux had faded well into the background noise. The subtractions
were performed with the {\sc hotpants}\footnote{http://www.astro.washington.edu/users/becker/v2.0/hotpants.html} routine.

\begin{figure}[htpb] % Figure 2
\epsscale{1.0}
\plotone{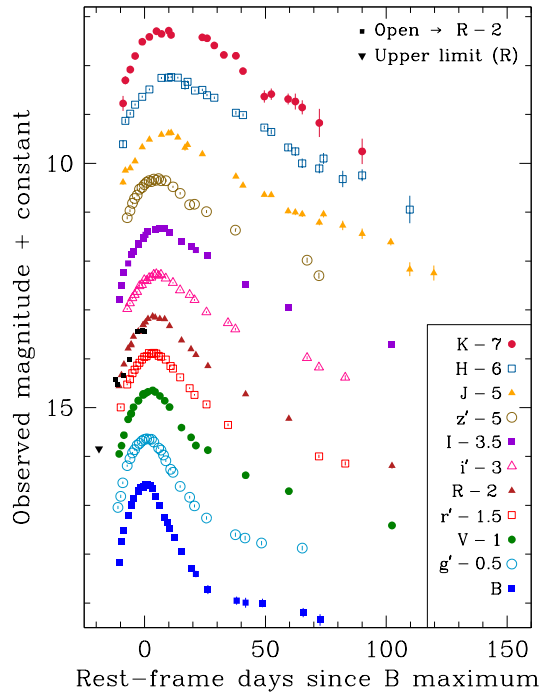}
\caption{$BVRIg'r'i'z'JHK$ light curves of SN2010as. Black squares
  correspond to open-filter photometry converted to $R$ band. The black
  triangle indicates the upper limit of the pre-discovery image, also
  converted from open filter to the $R$-band scale.
  \label{fig:lcs}}
\end{figure}

Photometry of the SN was performed with PSF fitting and 
calibrated relative to a local sequence of stars in the
field of NGC~6000. The field stars are indicated in
Figure~\ref{fig:fc}. They were 
selected to cover a wide range of brightnesses and colors, and to
cover the field around the SN while avoiding the bright parts of the
host galaxy. The photometric sequence was calibrated to the standard
Landolt \citep{landolt92} and Sloan \citep{smith02} systems using
observations from at least two photometric nights (most
typically, 3--5 nights). The resulting standard magnitudes of the
field stars are listed in Table~\ref{tab:lseq} for $BVRI$, and in
Table~\ref{tab:sseq} for $g'r'i'z'$. Aparent magnitudes were
  computed for the SN in the Landolt and Sloan systems, as listed in
  Tables~\ref{tab:lsn} and \ref{tab:ssn}, respectively. Resulting 
optical light curves are shown in Figure~\ref{fig:lcs}.

Near-infrared (NIR) imaging was obtained in the $JHK$ bands using the
Gamma-Ray Burst Optical and Near-Infrared Detector
\citep[GROND;][]{greiner08} installed on the MPI $2.2$~m telescope
at the La Silla Observatory, Chile. NIR observations started two days
after discovery and continued for about 130 days. The
field of view covered by the detector was of $10' \times 10'$, with a
pixel scale of $0{\farcs}6$. Reductions were carried out using
standard pyraf/IRAF\footnote{IRAF, the Image Reduction and Analysis
   Facility, is distributed by the National Optical Astronomy
   Observatory, which is operated by the Association of Universities
   for Research in Astronomy (AURA), Inc., under cooperative agreement
   with the National Science Foundation (NSF); see
   \url{http://iraf.noao.edu}.} routines to perform dark subtraction,
flat-fielding, sky subtraction, image alignment and stacking.  

Host-galaxy background flux was subtracted using deep images obtained
with GROND on 2011 May 20, that is nearly 430 days after discovery. PSF
photometry was performed relative to nine field stars in the Two
Micron All Sky Survey (2MASS), which are listed in Table~\ref{tab:nseq}.
Transformation equations to the standard system were adopted from
\citet{greiner08}. The resulting SN photometry in $JHK$ is listed in
Table~\ref{tab:nsn} and plotted in Figure~\ref{fig:lcs}.  

\subsection{Spectroscopy}
\label{sec:spec}

Intensive spectroscopic monitoring of SN~2010as was initiated by the
MCSS soon after the discovery, on 2010 March 20 UT, i.e. at $-10.5$
days relative to the time of $B$-band maximum light (see
Section~\ref{sec:lcs}). Seventeen epochs of spectroscopy were obtained,
ten of them before maximum light with a nearly nightly cadence. Six
post-maximum spectra were obtained on a regular basis for about four
months. An additional spectrum was obtained at ten months after maximum. 

Most of the spectra covered the optical range and were obtained using
long-slit, low-resolution spectrographs. The instruments used were the
Wide Field CCD Camera (WFCCD) mounted on the 2.5~m du Pont Telescope at Las
Campanas Observatory, the Gemini Multi-Object Spectrograph
\citep[GMOS;][]{hook04} on the 
Gemini South telescope, and the Goodman High Throughput Spectrograph
\citep[GHTS;][]{clemens04} on the Southern Astrophysical Research
(SOAR) Telescope. The final spectrum was obtained
using the Inamori Magellan Areal Camera and Spectrograph 
\citep[IMACS;][]{dressler11} on the Magellan Baade 6.5~m telescope at
Las Campanas Observatory. Medium-resolution, optical and NIR
spectroscopy was obtained on four epochs before maximum light using the
X-Shooter Spectrograph \citep{vernet11} mounted on the ESO Very Large
Telescope (VLT). A log of the spectroscopic observations is given in
Table~\ref{tab:speclog}. 

\begin{figure}[htpb]% Figure 3
\epsscale{1.0}
\plotone{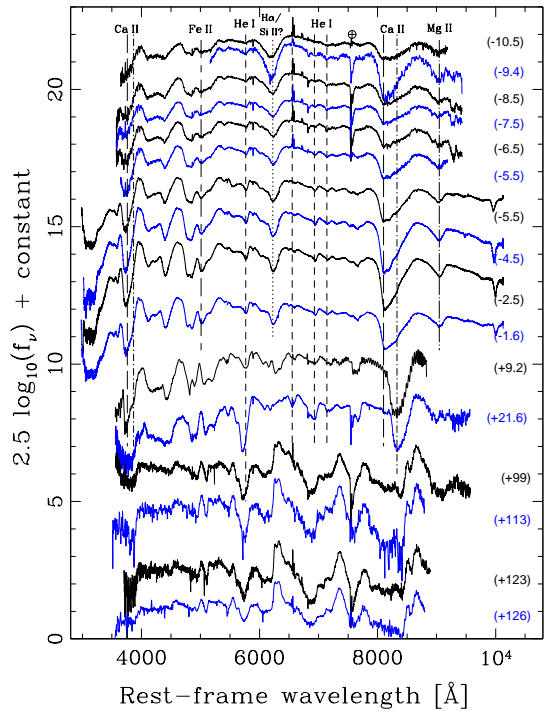}
\caption{Optical spectroscopic time-series of SN~2010as between $-11$ and
  $+126$ days with respect to $B$-band maximum light. The rest-frame
  epoch of each spectrum relative to $B$-band maximum is indicated in
  parentheses. Ions responsible for the main spectral features are
  labeled at the top. Dashed lines show the location of \ion{He}{1}
  lines blueshifted 
  by 5700 km s$^{-1}$. The dotted line indicates the location of
  H$\alpha$ at 15500 km s$^{-1}$. Dot-dashed and long-dashed lines
  show the two components of \ion{Ca}{2} lines at 6000 km s$^{-1}$ and
  14000 km s$^{-1}$, respectively. The dot-long-dashed line shows the
  location of \ion{Mg}{2}~$\lambda$9227 blueshifted by 6000 km
  s$^{-1}$. The short-long dashed line indicates the location of
  \ion{Fe}{2}~$\lambda$5169 at 9500 km s$^{-1}$.
  \label{fig:specopt}}
\end{figure}

\begin{figure}[htpb]% Figure 4
\epsscale{1.0}
\plotone{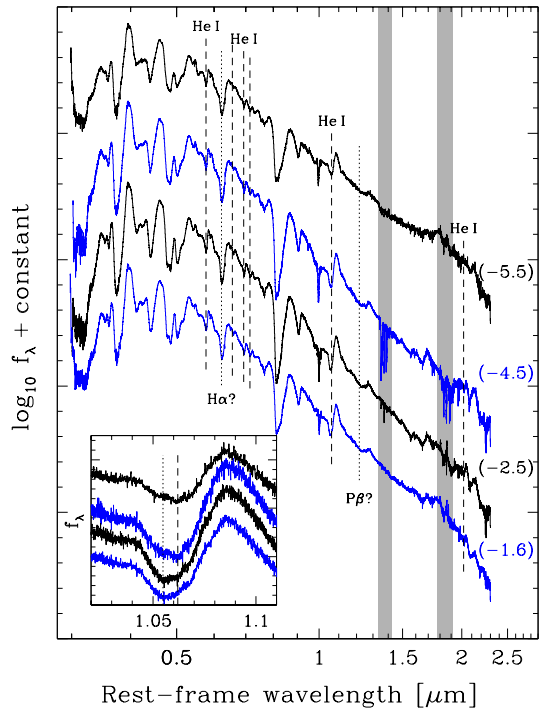}
\caption{X-Shooter spectra of SN~2010as obtained between $-5.5$ and
  $-1.6$ days with respect to $B$-band maximum light. The spectra
  simultaneously cover the optical and NIR ranges. The numbers in
  parentheses indicate the epoch of the spectra. Dashed vertical lines
  show the location of \ion{He}{1} lines blueshifted 
  by 5700 km s$^{-1}$ while dotted lines indicate the expected
  location of H$\alpha$ and P$\beta$ at 15500 km s$^{-1}$. The gray regions
  show the wavelength ranges where the atmospheric transmission drops
  roughly below 50\%. The inset shows a zoom around the
  \ion{He}{1}~$\lambda$10830 line (note the flux scale is
  linear). Vertical dashed and dotted lines show the location of the
  line at 5800 km s$^{-1}$ and 7800 km s$^{-1}$, respectively.
  \label{fig:specnir}}
\end{figure}

Reduction of the WFCCD, GHTS and IMACS spectra was performed using
standard IRAF routines. Dedicated pipelines were
employed for the GMOS and X-Shooter data. When available, spectra of
bright featureless stars observed close in time and zenith distance to
the SN spectrum were used to correct the telluric absorptions. Flux
calibration was performed using standard star spectra observed on each
night. The resulting wavelength- and flux-calibrated spectra are shown
in the optical range in
Figure~\ref{fig:specopt}. Figure~\ref{fig:specnir} shows the four
spectra from X-Shooter that cover the optical and NIR ranges.

\section{DATA ANALYSIS}
\label{sec:ana}

\subsection{Light Curves}
\label{sec:lcs}

\noindent We performed polynomial fits to the light curves of
SN~2010as in order to obtain time of maximum light and peak
brightness in all bands. In all cases we used polynomials of 3rd or
4th order. To avoid raising the polynomial order and
loosing accuracy at maximum light, we only considered data obtained
earlier than $\approx$30 days after maximum. The results of the fits
are listed in Table~\ref{tab:lcfits}. In this paper we adopt the
time of maximum light in $B$ band, $t_B({\mathrm{max}})=2455286.4$
(JD), as the reference epoch. We note an overall drift in the epoch of 
maximum light toward later times as we move to longer wavelength. This
behavior is commonly seen in SE SNe \citep[see, for example, the cases
  of SN~1999ex and SN~2007Y in][respectively]{stritzinger02,stritzinger09}. 
The rise and fall timescale of the light curves increases
systematically as the effective wavelength increases.

The optical photometry covers from $-11$ days to $\approx$100 days
relative to maximum light. The NIR coverage starts two days after and
extends until 120 days past maximum. After $\approx$$+70$ days
$B$-band and $g'$-band coverage stops and the coverage in other
optical bands becomes scarce. The NIR light curves continue with good
sampling but become noisier after about that epoch due to the
increasing relative background level. The $B$ and $g'$ light curves
show a flattening after about day $+30$, following a steep decline
from maximum light. Other bands show a more gradual, linear decline
extending until $\approx$$+100$ days.

\subsection{Color and Reddening}
\label{sec:col}

\begin{figure}[htpb]% Figure 5
\epsscale{1.0}
\plotone{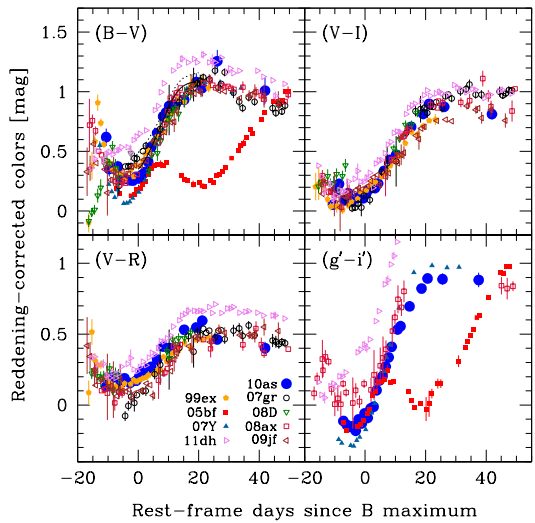}
\caption{Color curves of SN~2010as (black dots) compared with other SE
  SNe. Colors were corrected for total reddening (see text for
  details). The solid curve in the upper-left panel shows the locus of the
  intrinsic $(B-V)$ colors of SE SNe as derived from the CSP sample
  (Stritzinger et~al., in preparation),
  with dotted lines indicating the 1-$\sigma$ dispersion. Comparison
  SE SNe from the literature comprise the Type~IIb SN~2008ax
  \citep{pastorello08} and 2011dh \citep{ergon14}, the Type~Ib
  SN~2008D \citep{modjaz09} and SN~2009jf \citep{valenti11,sahu11},
  the Type~Ic SN~2007gr \citep{valenti08,hunter09}, and finally the transitional
  Type~Ib/c SN~1999ex \citep{stritzinger02}, SN~2005bf
  \citep{folatelli06} and SN~2007Y \citep{stritzinger09}.
  \label{fig:colora}}
\end{figure}

\noindent 
The location of SN~2010as very near the core of the starburst galaxy
NGC~6000 makes this object prone to substantial extinction from
dust in the interstellar medium of its host galaxy (see
Section~\ref{sec:env}). We thus used the 
observed SN colors to attempt to determine the amount of reddening and
thereby of extinction in any given band. To 
that end, we first corrected colors by the significant---albeit
well-known---Galactic extinction. According to the re-calibrated
infrared dust maps of \citet{schlafly11} the Galactic reddening in the
direction of NGC~6000 is $E(B-V)_{\mathrm{Gal}}=0.15$ mag. With this correction
applied, we compared the $(B-V)$ color with those of six 
reddening-free Type Ib and Ic SNe observed by the {\em Carnegie Supernova
Project} \citep[CSP][]{hamuy06}. These reddening-free SNe were
selected from the lack of narrow \ion{Na}{1}~D absorptions in the
spectra, and by being relatively away from star-forming regions in
their host galaxies. They were used to calculate an average intrinsic
$(B-V)$ color curve between the time of $B$-band maximum light 
and 20 days after, which is shown in the upper-left panel of
Figure~\ref{fig:colora} (Stritzinger et~al., in preparation). The
color curve evolves monotonically from 
$(B-V)=0.35$ mag to $(B-V)=1.08$ mag in such epoch range, with a
typical dispersion of $0.06$--$0.14$ mag. SN~2010as exhibits a similar
trend in the $(B-V)$ color evolution but with systematically redder
observed colors. By assuming the deviation is due to reddening in the
host galaxy, we derived $E(B-V)_{\mathrm{Host}}=0.42 \pm 0.02$ mag
from the weighted average of the deviations. A systematic uncertainty
of $\approx$$0.1$ mag can be added to this due to the dispersion in
the intrinsic colors.

We also compared our $(V-R)$ colors with the average intrinsic colors for
Type Ib and Ic SNe derived by \citet{drout11}. These averages are
taken at 10 days past the time of maximum in $V$ and in $R$ bands,
with $\langle(V-R)\rangle_{V10}=0.26 \pm 0.06$ mag and
$\langle(V-R)\rangle_{R10}=0.29 \pm 0.08$ mag, respectively. After
correcting for Galactic extinction, we obtain $(V-R)_{V10}=0.62\pm0.2$
mag and $(V-R)_{R10}=0.65\pm0.2$ mag for SN~2010as. Averaging the
differences with both reference values, we obtain
$E(V-R)_{\mathrm{Host}}=0.36 \pm 0.05$ mag. If we assume a standard
reddening law from \citet{cardelli89} with total-to-selective
absorption coefficient $R_V=3.1$, this can be
converted to $E(B-V)_{\mathrm{Host}}=0.67 \pm 0.09$ mag. This value is
in disagreement---although at a $1.8$-$\sigma$ level---with that
obtained from the comparison of $(B-V)$ 
colors of the CSP sample. The discrepancy may be due to the fact that
\citet{drout11} derived their intrinsic colors using a sample of
ten SE SNe from the literature, including Type Ib, Ic,
IIb and three Ic-BL objects. The photometric systems and estimated
extinctions affecting those SNe are thus of very diverse origin. Our
comparison of $(B-V)$ colors with those in the CSP sample is probably
more robust. Also, we consider this higher reddening to be
unrealistic because it would produce too blue colors in other bands,
as shown below.

We further estimated reddening from the Balmer decrement based on the
narrow emission-line flux ratio of H$\alpha$ to H$\beta$. This was done on
spectra extracted nearby the SN location and adopting case B
recombination with $n_e \sim 10^2$--$10^4$ cm$^{-3}$ and $T_e=10^4$
K \citep{osterbrock89}. From this we obtained a value of  
$E(B-V)_{\mathrm{Host}}=0.44$ mag, in agreement with the value from
the $(B-V)$ color. In addition, we could clearly detect three sets of
\ion{Na}{1}~D absorption lines in the spectra at all phases. One
component appeared at rest, in agreement with the existence of
Milky-Way gas and dust in the line of sight, and the other two
components appeared with recession velocities of about 2020 and 2240
km s$^{-1}$, i.e. at $\approx$$-170$ and $\approx$50 km s$^{-1}$ with
respect to the systemic recession velocity of the host galaxy. 
We measured the equivalent widths (EW) of the \ion{Na}{1}~D doublet
and found no evidence of variation with phase. The average EW was
of $0.7$ \AA\ for the Galactic component, and $2.1$ \AA\ and $0.5$
\AA\ for the two components in NGC~6000. Although the EW of
\ion{Na}{1}~D does not provide an accurate estimate of the color
excess \citep{phillips13}, the relative strength of the
absorptions is in agreement with the presence of higher extinction in
the host galaxy, as compared with that in the Milky Way. 

\begin{figure}[htpb]% Figure 6
\epsscale{1.0}
\plotone{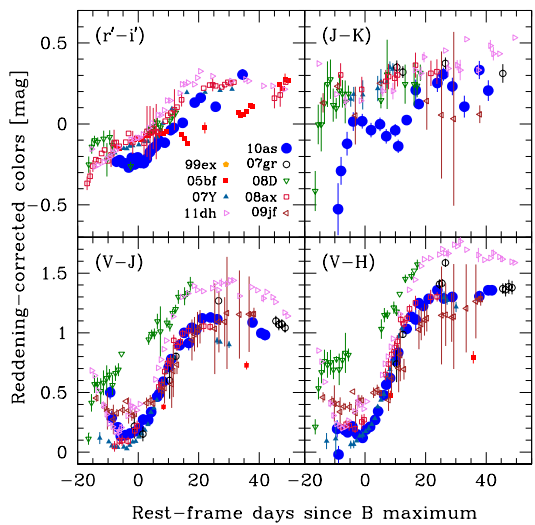}
\caption{Color curves of SN~2010as (black dots) compared with other SE
  SNe. Colors were corrected for total reddening (see text for
  details). See references of data sources in Figure~\ref{fig:colora}.
  \label{fig:colorb}}
\end{figure}

Based on the analysis above, we adopted $E(B-V)_{\mathrm{Host}}=0.42 \pm
0.1$ mag to compute the reddening-corrected colors. Selected color
indices are shown in Figures~\ref{fig:colora} and \ref{fig:colorb} in
comparison with intrinsic colors for a sample of well-observed SE
SNe. For the comparison SNe, colors were corrected for Galactic and
host-galaxy reddening by adopting color-excess values from the
literature and a standard reddening law with $R_V=3.1$. As expected,
the corrected $(B-V)$ colors of SN~2010as 
(upper-left panel of Figure~\ref{fig:colora}) agree well with most SE
SNe. However, using $R_V=3.1$ for the host-galaxy reddening of
SN~2010as, its other colors, especially those involving a wide wavelength
span, appeared too blue in comparison with the rest of the
objects. For example, SN~2010as would be intrinsically bluer than all
other SE SNe near maximum light by $\approx$$0.2$ mag in $(V-I)$ and
$(r'-i')$, by $\approx$$0.4$ mag in $(V-J)$, and by $\approx$$0.6$ mag
in $(V-H)$. We attributed this behavior to the adopted value of $R_V$
for the host-galaxy reddening. By lowering it to $R_V=1.5$, as shown
in the figures, we obtained a much better agreement with the
comparison sample, consistently for all the colors. Similarly low
values of $R_V$ have typically been found for Type~Ia SNe with significant 
reddening \citep[see][]{phillips13}, and in studies of Type~II-Plateau
SNe \citep{olivares10}. The origin of low $R_V$ values
may be linked to the properties of the dust grains, or to the
presence of a dense circumstellar shell of dust around the SN. In
Section~\ref{sec:lbol} we present further support for a low $R_V$ from
our calculation of the bolometric luminosity. We thus
adopted a non-standard, low, reddening law parameter of $R_V=1.5$ for
SN~2010as. Due to the observed dispersion of colors among SE SNe,
which is significantly larger than that found among Type~Ia 
SNe, a precise derivation of $R_V$ is not possible. In the future,
such a study may be addressed provided there is a consistent
photometric dataset for a homogeneous SN sample.

From Figures~\ref{fig:colora} and \ref{fig:colorb} we see that there
is a group of SE SNe that agree in their color evolution, especially
from a few days before maximum and on. In the optical regime colors
tend to agree better whereas when near-infrared bands are involved
(see Figure~\ref{fig:colorb}) the dispersion becomes large, e.g., as
much as 1 mag in $(V-J)$ and $(V-H)$. The outlying object in this
sample is the peculiar SN~2005bf that showed doubly peaked light
curves and very large luminosity
\citep{anupama05,tominaga05,folatelli06}. Its colors 
are consistent with most SE SNe near maximum light\footnote{For
  SN~2005bf, we adopted the time of the first peak in the
  $B$ band \citep{folatelli06}.}, but as the SN evolved to its second peak,
optical colors became blue again and remained so for about 40 days,
drastically deviating from the rest of the SNe.  

We computed reddening-free peak absolute magnitudes in all observed
bands using the reddening estimate derived above and the distance to the
host galaxy of $27.38 \pm 4.74$ Mpc (distance modulus of
$\mu = 32.16 \pm 0.36$ mag) from the NASA/IPAC Extragalactic
Database (NED). This distance is an average of six
  measurements based on the Tully-Fisher method from the Nearby Galaxy
  Catalog (Tully,~R.~B.\ 1988), and later by
  \citet{terry02,theureau07}. The resulting absolute magnitudes are
listed in Table~\ref{tab:lcfits}.

\subsection{Bolometric Luminosity}
\label{sec:lbol}

\noindent We used the broad-band photometry to compute the integrated
flux, $F_{B\rightarrow K}$, in the range between $B$ and $K$
bands. For this purpose, we converted observed 
magnitudes, corrected by extinction as in Section~\ref{sec:col}, to
monochromatic fluxes at the effective wavelength of each filter. The
conversions of magnitude to flux were performed using the spectral
energy distribution of Vega as given by \citet{bohlin04a}. 
When an observation in a given band was missing, we used a
low-order interpolation of the neighboring observations in that
band. Monochromatic fluxes at a given date were then trapezium-integrated in 
wavelength to produce the pseudo-bolometric flux $F_{B\rightarrow K}$. 

\begin{figure}[htpb]% Figure 7
\epsscale{1.0}
\plotone{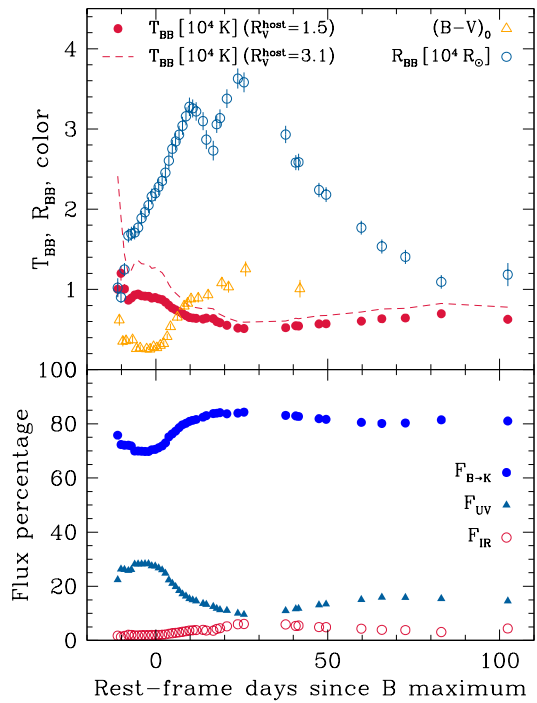}
\caption{{\em Top:} Evolution of the fitted BB temperature ({\em
    filled circles}) and radius ({\em open circles}), and of the
  intrinsic $(B-V)$ color ({\em triangles}), as a function of the
  rest-frame time since maximum bolometric luminosity. {\em Bottom:}
  Estimated 
  contributions to the total flux outside the observed wavelength
  range. $F_{\mathrm{UV}}$ is the estimated flux blueward of the $B$
  band, and $F_{\mathrm{IR}}$ is the estimated flux redward of the $K$
  band (see text for details).
  \label{fig:bbfits}}
\end{figure}

In order to account for the missing flux outside the wavelength range
covered by our photometry, we used extrapolations. On the UV regime we
assumed that the flux dropped linearly from the value measured at the
effective wavelength of the $B$ band (at 447 nm) to zero flux at 200
nm and that the flux blueward of that wavelength was negligible
\citep{bersten09,lyman14}. This assumption is based on observations of 
core-collapse SNe \citep{panagia03,bufano09}. We simply integrate the
flux in such triangle to obtain the UV contribution, $F_{\mathrm{UV}}$. The
flux redward of the $K$ band, $F_{\mathrm{IR}}$, was estimated by
extrapolating a black-body (BB) fit of the observed monochromatic
fluxes. Table~\ref{tab:lbol} provides the fitted BB radii,
$R_{\mathrm{BB}}$, and temperatures, $T_{\mathrm{BB}}$. If the fitted
BB temperature was high enough, the Rayleigh-Jeans (R-J) approximation was
adopted\footnote{Adopting the R-J condition of 
  $hc/\lambda kT \leq 0.5$, for the effective wavelength of the $K$
  band, we obtain $T_{\mathrm{BB}}\geq 1.32 \times 
  10^{4}$ K.}. Otherwise, the BB was integrated between the effective
wavelength of the $K$ band and 15000 nm, which produces a good
  approximation of the integration until infinite wavelength. The
total bolometric flux was thus computed as
$F_{\mathrm{Bol}}=F_{B\rightarrow
  K}+F_{\mathrm{UV}}+F_{\mathrm{IR}}$. Figure~\ref{fig:bbfits} shows
the percentage contribution of $F_{B\rightarrow K}$,
$F_{\mathrm{UV}}$, and $F_{\mathrm{IR}}$ with time, along with 
the evolution of the BB temperature and radius, and the
reddening-free $(B-V)$ color. Due to the long wavelength 
coverage of the data, the estimated contribution of the extrapolated
fluxes is moderate at all times. The actually measured flux,
$F_{B\rightarrow K}$, is always greater than 70 percent of the total
flux. As shown in Figure~\ref{fig:bbfits}, before maximum light the 
ejecta are hot and the UV contribution is relatively
important. Nevertheless it is always below 30\%. The IR contribution
is small at all times and remains below 6\% during the time considered
here.  In Figure~\ref{fig:bbfits} we also show the BB temperature evolution
when assuming a host-galaxy reddening law with $R_V=3.1$. Initially,
the resulting temperatures are unrealistically high. This is an
additional indication of a low $R_V$, as derived in Section~\ref{sec:col}.

\begin{figure}[htpb]% Figure 8
\epsscale{1.0}
\plotone{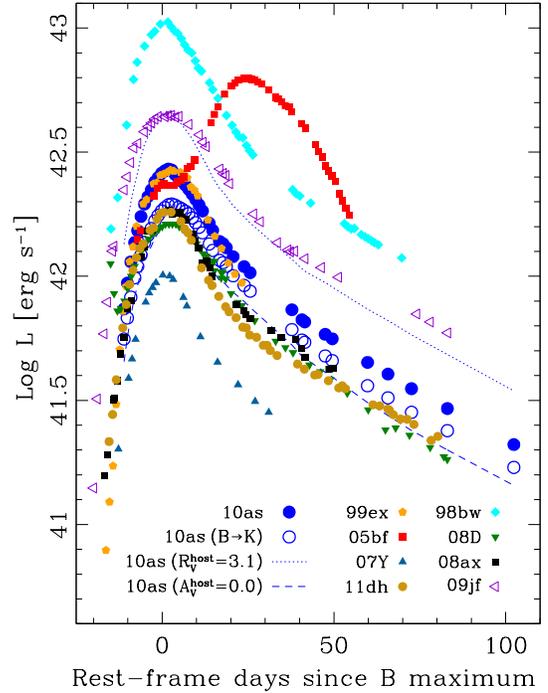}
\caption{Bolometric light curve of SN~2010as compared with SE SNe from
  the literature (see references in the text). Filled blue circles show the
  estimated bolometric luminosity of SN~2010as, after correcting for
  total extinction, with $R_V=1.5$ for the host-galaxy
  extinction. Open blue circles show the contribution that was 
  actually observed, integrated between the $B$ and $K$ bands. The
  dotted curve indicate the bolometric luminosity that would result by
  assuming $R_V=3.1$ for the host-galaxy reddening. The dashed curve
  shows the case of no extragalactic extinction. Comparison SNe are:
  the transitional Type~Ib/c SN~1999ex \citep{stritzinger02}, 
  SN~2005bf \citep{folatelli06}, and SN~2007Y \citep{stritzinger09}; the
  Type~Ib SN~2008D \citep{modjaz09} and SN~2009jf
  \citep{sahu11,valenti11}; the Type~IIb SN~2008ax
  \citep{pastorello08,taubenberger11} and SN~2011dh \citep{ergon14}; and
  the broad-line Type~Ic SN~1998bw \citep{clocchiatti11}.
  \label{fig:lbolcomp}}
\end{figure}

Bolometric fluxes were converted to luminosities adopting a distance
of $27.38 \pm 4.74$ Mpc from NED. Table~\ref{tab:lbol} lists the
luminosity integrated between $B$ and $K$ ($L_{B\rightarrow K}$) and the
estimated bolometric luminosity ($L_{\mathrm{Bol}}$). In Table~\ref{tab:lcfits}
we included the peak apparent and absolute bolometric magnitudes obtained
from a polynomial fit to the bolometric light curve. The bolometric light
curve is shown in Figure~\ref{fig:lbolcomp} along with several
well-studied SE SNe. For comparison, we also show
$L_{B\rightarrow K}$ computed as described above, and the resulting
$L_{\mathrm{Bol}}$ for zero host-galaxy reddening, and for
$E(B-V)_{\mathrm{host}}=0.42$ mag with $R_V=3.1$ (dashed and dotted
lines, respectively). For SNe~2005bf
and 2008D we obtained the bolometric luminosities directly from the
literature because they were computed in a similar fashion as that
described here. For the rest of the objects we
calculated bolometric luminosities using the color-based bolometric
corrections provided in Table~2 of \citet{lyman14}. All available
colors were employed for each SN, and the resulting luminosities were
averaged. Estimates of Galactic extinction and distances were obtained
from NED. Host-galaxy extinction values were adopted from the
literature. 

The peak bolometric luminosity of SN~2010as occurred about $1.5$ days
after the peak in the $B$ band, with a value of $(2.7 \pm 1.0) \times
10^{42}$ erg s$^{-1}$. The uncertainty in the luminosity was estimated
by considering uncertainties in distance, extinction and the estimated
error in the extrapolations done to compute $F_{\mathrm{UV}}$, and
$F_{\mathrm{IR}}$. The uncertainty in distance implies a 34\%
uncertainty in luminosity. This is the dominant component, and is
constant in time. From the $0.1$ mag uncertainty in $E(B-V)$, we
derived an uncertainty of $\approx$10\% in luminosity. We assumed the
extrapolations to contain a 20\% uncertainty, which corresponds to 6\%
of the total flux. Assuming the explosion happened at the midpoint 
between the last 
non-detection and the discovery, the rise time to maximum luminosity
was of $16.5 \pm 3$ days. The bolometric light curve of SN~2010as is
similar to that of SN~1999ex. The first peak of SN~2005bf also shows a
similar evolution, before it re-brightens to a wide and luminous main
peak. The Type~IIb SNe~2008ax and 2011dh show similar light-curve
shapes with slightly slower rise and lower peak luminosity. The
Type~Ib objects shown here (SNe~2008D and 2009jf) have wide
bolometric light curves---with an initial decline in the case of
SN~2008D.

\subsection{Spectroscopic Properties}
\label{sec:spep}

\noindent 
The spectroscopic follow-up of SN~2010as continued between $-11$ and
$+309$ days relative to $B$-band maximum light, with nearly daily
coverage before maximum. This allowed us to perform a detailed study
of the evolution of spectral properties for this object. 

\subsubsection{Photospheric Phase}
\label{sec:specmax}

We study here the spectra obtained between $-11$ and $+22$ days
relative to $B$-band maximum light, when a clear continuum emission
could be identified. As seen in
Figure~\ref{fig:specopt}, the first spectrum has mostly Type~Ic
characteristics. It is dominated by wide \ion{Ca}{2} and \ion{Fe}{2}
absorption features with very slight signs of \ion{He}{1}
lines. Helium lines slowly increase in strength with
time and become conspicuous in the optical range not earlier than a
few days before maximum. \ion{He}{1}~$\lambda$10830 is clearly
identifiable in our 
pre-maximum NIR spectra (Figure~\ref{fig:specnir}), as is expected
since this is usually the strongest helium feature. \ion{Ca}{2} and
\ion{Fe}{2} absorptions remain 
strong throughout the epoch range sampled here. In particular, \ion{Ca}{2}
lines evolve from being dominated by large Doppler-shift components
before maximum, to much lower shifts after maximum.
A strong absorption at about 6200 \AA\ is also present at least until 
maximum light. Such feature has been observed in
other Type~Ib and Ic objects and its identification is unclear
\citep{branch06,elmhamdi06,ketchum08,spencer10}. 
Here we will evaluate the most probable associations, which are
\ion{Si}{2}~$\lambda$6355 or H$\alpha$. The
identification of this feature is crucial for the
understanding of the nature of SN~2010as in the context of
SE SNe. 

\begin{figure}[htpb]% Figure 9
\epsscale{1.0}
\plotone{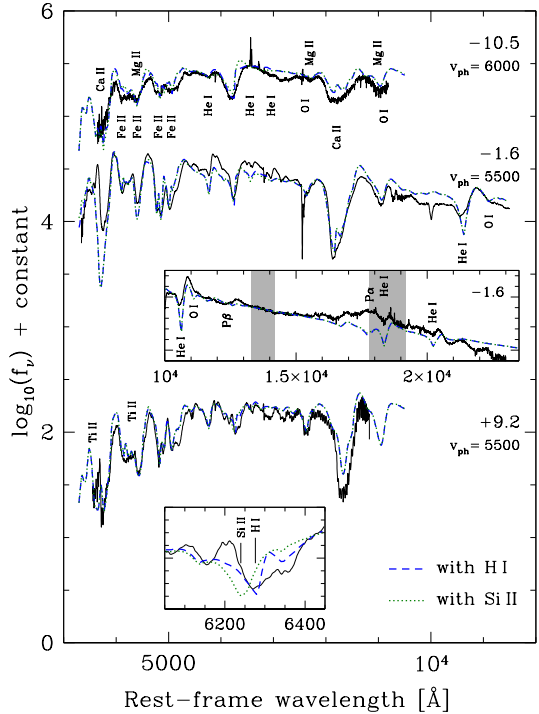}
\caption{Synthetic spectra computed with SYNOW compared with the
  spectra of SN~2010as obtained at $-10.5$, $-1.6$ and $+9.2$ days
  (solid black lines). The SYNOW spectra including \ion{H}{1} at high
  velocity are shown with dashed blue lines. Alternative calculations without
  hydrogen and with \ion{Si}{2} are shown with dotted green
  lines. The main ions that contribute to the observed features are
  indicated. The top inset graph shows the NIR range of the spectrum
  from $-1.6$ days. The bottom inset panel shows the detail around the
  H$\alpha$ line at $+9.2$ days where we indicated the location of the
  \ion{Si}{2}~$\lambda$6355 line at photospheric velocity, and the
  H$\alpha$ line detached at about 10000 km s$^{-1}$. 
  \label{fig:synow}}
\end{figure}

In order to provide a more secure identification of the elements 
present in the ejecta and their approximate distribution, we 
performed synthetic spectrum calculations using the SYNOW code
\citep[see][and references therein]{branch02}. Figure~\ref{fig:synow}
shows SYNOW calculations compared with 
spectra observed at three different epochs. In this simple model, the
continuum emission is assumed to be due to a black body of given
temperature, $T_{\mathrm{BB}}$, defined at a sharp photosphere. The
ejecta distribution is given in velocity space, which can be
interpreted as a spatial coordinate by assuming homologous
expansion. Line profiles are computed in the Sobolev approximation by
arbitrarily fixing a distribution of optical depth ($\tau$) as a
function of velocity for each ionic species. The photospheric velocity,
$v_{\mathrm{ph}}$, thus fixes the minimum shift of absorption lines
in the spectrum. Relative line strengths for each species are
determined by fixing its excitation temperature, $T_{\mathrm{exc}}$,
and assuming local 
thermal equilibrium. Here we adopted consistent $T_{\mathrm{exc}}$ of
8000--10000 K for all species, except for \ion{H}{1} as explained
below. Considering the simplifications of the model, 
the synthetic spectra agree quite well with the observations. 

The main species included in the synthetic spectra were \ion{He}{1},
\ion{O}{1}, \ion{Mg}{2}, \ion{Ca}{2}, \ion{Ti}{2} and
\ion{Fe}{2}. Additional presence of \ion{Na}{1} is probable after maximum
light. The inclusion of \ion{H}{1} or \ion{Si}{2} will be  
discussed below. In order to reproduce the small blueshift of most
observed lines, we needed to set low values of $v_{\mathrm{ph}}$ even
at the earliest 
phases. For the first spectrum at $-10.5$ days we adopted
$v_{\mathrm{ph}}=6000$ km s$^{-1}$, and this value remained nearly
unchanged until nine days after maximum light. Only after that
time did the photospheric velocity decrease to $\approx$4000 km
s$^{-1}$ at $+22$ days, and $\approx$3500 km s$^{-1}$ at about $+100$
days. For all the elements listed above the line-forming region was
consistent with being attached to the photosphere. Even at such low
velocities, the line profiles appeared broad, which required a
shallow dependence of the optical depth as a function of velocity. For
all the species we adopted a power law for $\tau \propto
v^{-\alpha}$, with $\alpha$ being 3 until maximum light and 4 or 5
after maximum. The assumed maximum optical depths of all elements 
grew consistently until nine days after maximum light.

The large blue extent of the \ion{Ca}{2} absorptions 
and the shape of the \ion{Fe}{2} complex at $\approx$5000 \AA\ required
extra components at high velocity. The calcium and iron line profiles 
were best reproduced by assuming an additional shell with a Gaussian
optical-depth distribution. The central velocity of the shells
decreased with time from 
11000 km s$^{-1}$ at $-10.5$ days to 7000 km s$^{-1}$ and 10000 km
s$^{-1}$ at maximum light for \ion{Ca}{2} and \ion{Fe}{2},
respectively. The \ion{Ca}{2} shell had a Gaussian width of  
$\sigma_v=5000$--6000 km s$^{-1}$, while a narrower \ion{Fe}{2} shell of
$\sigma_v=2000$ km s$^{-1}$ was invoked. Both high-velocity components had
disappeared by nine days after maximum.

The strong absorption at 6200 \AA\ was not accounted for by any of the
species described above. We studied whether it was reproduced by
\ion{Si}{2}~$\lambda$6355 forming near the photosphere, by
high-velocity (detached) H$\alpha$, or by a combination of both.
As shown in Figure~\ref{fig:synow}, both the synthetic 
\ion{Si}{2} and H$\alpha$ lines provide reasonable approximations to the
observations before and at maximum light. Other lines of the same
species are too weak to be distinguished, as indicated by the fact
that both synthetic spectra are identical. An exception to this may be
P$\alpha$ which, depending on the adopted excitation temperature, can
become quite strong in the SYNOW spectrum. However, by adopting
$T_{\mathrm{exc}}\approx 5000$ K, the strength of P$\alpha$ could be
reduced to match the observation, as shown in the upper inset of
Figure~\ref{fig:synow}. We note that P$\alpha$ may be partly affected
by atmospheric absorption. While P$\beta$ is not noticeable in the
synthetic spectrum, the data show an absorption matching this line
blueshifted by the same velocity as H$\alpha$. This line is present in
all four of our NIR spectra (see Figure~\ref{fig:specnir}). Moreover, the
presence of hydrogen is further supported by the spectrum at $+9$ days
(see lower inset in Figure~\ref{fig:synow}).
In that spectrum the \ion{Si}{2}~$\lambda$6355 line fails to 
reproduce the location of the observed absorption that is still
identifiable among \ion{Fe}{2} lines. Instead, H$\alpha$ shifted by about
10000 km s$^{-1}$ produces a good agreement with the observation.
{\em We can conclude from the SYNOW analysis that SN~2010as showed at least
a small amount of hydrogen present at high velocity.}

\begin{figure}[htpb]% Figure 10
\epsscale{1.0}
\plotone{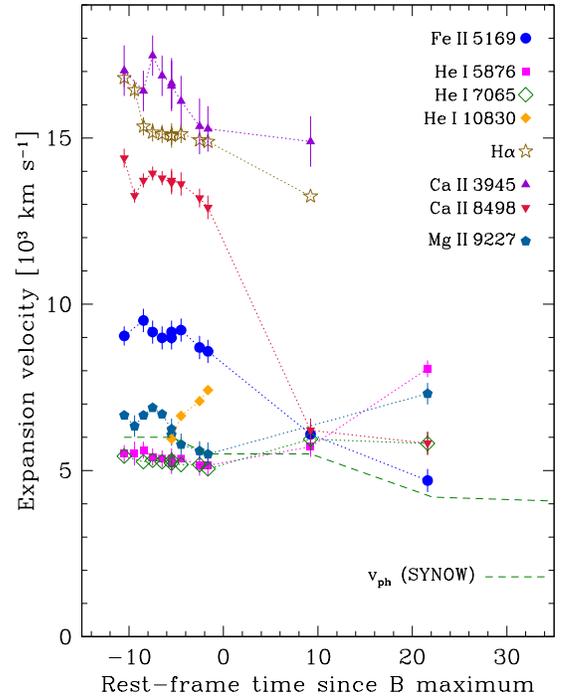}
\caption{Expansion velocities of SN~2010as as
  measured from the blueshift of the maximum absorption on
  well-identified lines. The dashed curve shows the photospheric
  velocity adopted in the SYNOW models. 
  \label{fig:veloc}}
\end{figure}

As we noted above from the SYNOW calculations, SN~2010as showed quite
low expansion velocities even at the earliest time. Moreover, the
photospheric velocity remained
nearly constant between $-11$ and $+9$ days. We studied this behavior
in greater detail by looking at the expansion velocities measured from
the shifts of the absorption minimum of the lines. Such velocities are
shown in Figure~\ref{fig:veloc} for \ion{He}{1}, \ion{Mg}{2},
\ion{Ca}{2} and \ion{Fe}{2} lines. The case of
H$\alpha$ is also shown. Before maximum light, the
lowest velocities correspond to \ion{He}{1} lines at $\approx$5500 km
s$^{-1}$, followed closely by \ion{Mg}{2}. The velocity of
\ion{He}{1}~$\lambda$10830 shows a rapid increase during the interval
sampled by the NIR observations. The inset of Figure~\ref{fig:specnir}
shows that this line appears as a double absorption with components at
$\approx$5800 km s$^{-1}$ and $\approx$7800 km s$^{-1}$, if both are
attributed to \ion{He}{1}. The measured velocity increases as the blue
component begins to dominate over the red one. It can also be that the
\ion{He}{1} line produces the red component at a constant velocity of
$\approx$5800 km s$^{-1}$, while the other absorption is due to
\ion{C}{1}~$\lambda$10693, although in that case the carbon velocity
would be too low ($\approx$4000 km s$^{-1}$), and no other carbon
lines are apparent in the spectra. Our SYNOW analysis
shows that the low expansion velocities is not a peculiarity of the
helium lines, but rather a property of the photosphere.

High-velocity components of \ion{Fe}{2} and \ion{Ca}{2} 
dominate before maximum, and so the measured velocities for these
species are significantly larger than the photospheric
one. \ion{Fe}{2}~$\lambda$5169 appears blueshifted at $\approx$9000 km
s$^{-1}$. This is a consequence of the blend of low- and high-velocity
components mentioned in our SYNOW analysis. \ion{Ca}{2} lines
and H$\alpha$ appear at 14000--17000 km s$^{-1}$. 
At nine days after maximum, all species show similar, low velocities
around 6000 km s$^{-1}$, including \ion{Fe}{2} and \ion{Ca}{2}, whose
high-velocity components disappear. By day $+22$ we notice a larger
dispersion of line velocities, with \ion{Fe}{2} showing the
lowest value. At this phase the \ion{He}{1}~$\lambda$5876 and $\lambda$7065
lines have velocities that differ by over 2000 km s$^{-1}$. This may
be due to the presence of \ion{Na}{1} blended with
\ion{He}{1}~$\lambda$5876, as suggested by the SYNOW calculation for
this epoch.  

\subsubsection{Transitional and Nebular Phases}
\label{sec:specneb}

\begin{figure}[htpb]% Figure 11
\epsscale{1.0}
\plotone{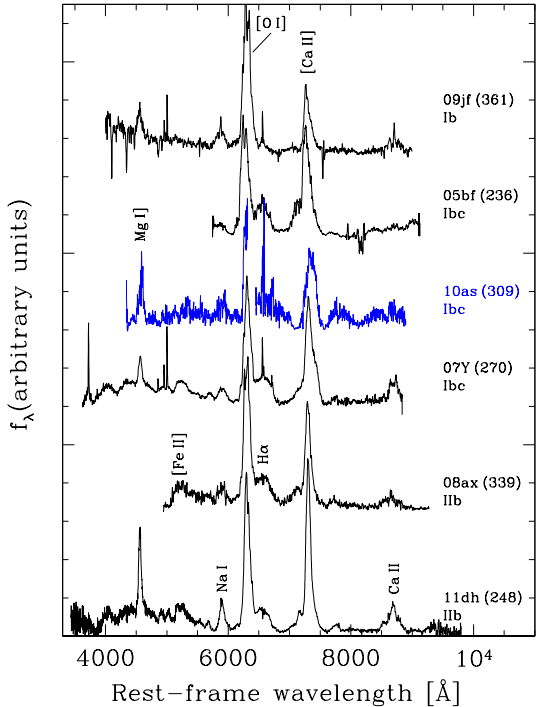}
\caption{Nebular spectrum of SN~2010as compared with those of other SE
  SNe at similar epochs (given in parentheses with respect to $B$-band
  maximum light). Ions for the main observed features are
  indicated. H$\alpha$ can be seen as a bump on the red side of the
  [\ion{O}{1}]~$\lambda$$\lambda$6300,6363 emission, except for the
  Type~Ib SN~2009jf. For SN~2010as, the cut in the wavelength coverage
  and the noise in the data make the identification of H$\alpha$
  difficult. 
  \label{fig:spcompneb}}
\end{figure}

Several spectra obtained between $+98$ and $+126$ days show the
transition to the nebular phase (see Figure~\ref{fig:specopt}), when
the continuum level decreases and emission lines begin to dominate. At
this stage the most prominent spectral features are due to oxygen and
calcium. \ion{O}{1}~$\lambda$7774 and the \ion{Ca}{2} infrared triplet
are seen predominantly in emission. Our last spectrum obtained at
$+309$ days is well in the nebular phase. It is shown in
Figure~\ref{fig:spcompneb} in comparison with those of other SE SNe.
At this stage the density has decreased enough for
forbidden lines to dominate. The most prominent of those lines are the
doublets of [\ion{O}{1}]~$\lambda\lambda$6300, 6363 and
[\ion{Ca}{2}]~$\lambda\lambda$7291, 7324. Apart from them, only
\ion{Mg}{1}$]$~$\lambda$4571 and possibly very weak
\ion{Na}{1}~D and \ion{Ca}{2} infrared triplet can be seen at this
late time. 

During the transitional phase, about 100 days after maximum, a bump
is observed on the red wing of the [\ion{O}{1}]~$\lambda\lambda$6300,
6363 emission. This can be seen in Figure~\ref{fig:specopt}, around the
narrow H$\alpha$ and [\ion{N}{2}] lines produced by an underlying
\ion{H}{2} region. Such feature has been observed in Type~IIb SNe and
in some Type~Ib objects and has been attributed to H$\alpha$. Such
emission is clearly visible in SNe~2005bf and 2007Y (see
Figure~\ref{fig:spcompneb}). In the
case of SN~2010as we observe that around $+100$ days its profile is
complex, which may indicate the presence of other lines. The
emission may still be present in the spectrum obtained at $+309$ days,
although the data are very noisy. We thus cannot provide further
evidence of the presence of hydrogen from the nebular spectra. 

\begin{figure}[htpb]% Figure 12
\epsscale{1.0}
\plotone{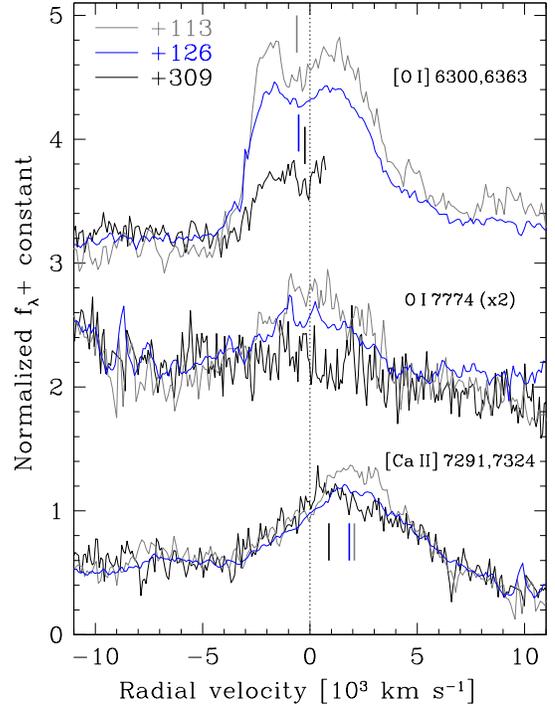}
\caption{The main emission lines observed in the transitional and
  nebular spectra of SN~2010as, plotted in velocity space. The spectra
  are from $+113$ days (gray), $+126$ days (blue), and $+309$ days
  (back). Reference wavelengths were assumed as 6300 \AA\ for the 
  [\ion{O}{1}]~$\lambda$$\lambda$6300,6363 doublet, 7774 \AA\ for
  \ion{O}{1}~$\lambda$7774, and 7307.5 \AA\ for 
  [\ion{Ca}{2}]~$\lambda$$\lambda$7291,7324.
  \label{fig:neblines}}
\end{figure}

In Figure~\ref{fig:neblines} we show the evolution of
[\ion{O}{1}]~$\lambda\lambda$6300, 6363,
[\ion{Ca}{2}]~$\lambda\lambda$7291, 7324, and \ion{O}{1}~$\lambda$7774
during the transition to the nebular phase. The strength of both the
permitted and forbidden \ion{O}{1} lines decrease substantially at
$+309$ days while the [\ion{Ca}{2}] feature remains nearly
constant. The [\ion{O}{1}] feature presents a double peak with a
central trough that appears slightly blueshifted relative to the
adopted reference wavelength of 6300 \AA. The other two features show
singly-peaked profiles, possibly with a central cusp in the case of
[\ion{Ca}{2}]. Such doubly-peaked [\ion{O}{1}] profiles have been
observed in other SNe
\citep{mazzali05,maeda08,modjaz08,taubenberger09} and their nature is
under debate. While \citet{maeda08} proposed that the shape is due to an 
asymmetric distribution of the inner oxygen-rich material,
\citet{milisavljevic10} suggest that at least in some SNe a
geometrical explanation is not necessary because we are simply
observing the two lines of the doublet. The latter scenario,
however, has problems in explaining deviations from the 3:1 flux ratio
expected between the $\lambda$6300 and $\lambda$6363 lines, and
the fact that a significant fraction of objects show a singly peaked
[\ion{O}{1}] profile \citep{maeda08,taubenberger09}. Finally,
\citet{maurer10} suggest that the central trough may be due to
high-velocity H$\alpha$ absorption. In this case, the velocity of
H$\alpha$ would be 13000 km s$^{-1}$, which is compatible with the
velocity measured nine days after maximum light (see
Figure~\ref{fig:veloc}).  
 
Assuming we are observing two components of the [\ion{O}{1}] doublet,
Figure~\ref{fig:neblines} shows a blueshifted and a redshifted peak,
which is in general agreement with the picture of an edge-on disk or
torus of O-rich material. The trough between the peaks appears
slightly blueshifted, which may be due to self-absorption in the
torus/disk structure that blocks part of the emission from the
receding sector that is farthest from the observer. We note that the
blueshift of the central trough decreases from $\approx$600 km
s$^{-1}$ at about 100 days to $\approx$200 km s$^{-1}$ at $+309$
days. This observation is consistent with a reduction in the internal
opacity as the ejecta expands. Interestingly, during this period the
[\ion{Ca}{2}]~$\lambda\lambda$7291, 7324 feature appears redshifted
from the adopted reference wavelength of $7307.5$ \AA. The amount of
redshift also decreases with time, from $\approx$2000 km
s$^{-1}$ to $\approx$900 km s$^{-1}$. This seems to be a peculiar
feature of SN~2010as, as in general 
the [\ion{Ca}{2}] emission appears at rest or slightly blueshifted in
other well-observed SNe. The origin of the redshift may be an
asymmetric distribution of the inner calcium-rich material, or a large
degree of contamination of this feature by lines emitted redward of
the central [\ion{Ca}{2}] wavelength. If such contamination were due
to [\ion{Fe}{2}], the lines around this [\ion{Ca}{2}] feature would
likely be stronger on the blue side ($\lambda\lambda$7155, 7172) than
on the red side \citep[$\lambda\lambda$7388, 7452;][]{garstang62}, and
thus it would not produce a net redshift. Although very noisy, the
\ion{Mg}{1}$]$~$\lambda$4571 line observed at $+309$ days also
appears redshifted by about 1400 km s$^{-1}$, which provides support
to the peculiar redshift of the [\ion{Ca}{2}] doublet. 

\section{ENVIRONMENT AND PROGENITOR OF SN~2010as}
\label{sec:prog}

\subsection{Host Galaxy and Environment}
\label{sec:env}

\noindent SN~2010as appeared very near the nucleus of the 
starburst galaxy NGC~6000, of morphological type ``SB(s)bc:'' 
\citep[RC3;][]{devaucouleurs91}. NGC~6000 
has been cataloged as a luminous infrared galaxy \citep[LIRG;][]{sanders03} 
and it is also luminous in the optical and the NIR ranges.
It has been reported as an active galaxy 
with nuclear a \ion{H}{2} region \citep{moran96}. 
By modeling mid-infrared observations,
\citet{siebenmorgen99} and \citet{siebenmorgen04} concluded that large
amounts of dust are present in the nuclear region of the galaxy, with
a central extinction between $A_V=20$ and $A_V=29$ mag. Adopting a distance 
of $27.38 \pm 4.74$ Mpc (NED), the SN location is at a projected
distance of only $490 \pm 23$ 
pc from the galaxy center. We therefore expect the SN to be located in
a region of rich star formation, with enriched gas and probably large
amounts of dust. Details about the underlying stellar
population at the specific location of the SN were obtained from the
{\em HST} pre-SN images and are described in Section~\ref{sec:presn}. 

We studied the abundance of metals in the vicinity of the SN and in
the galaxy as a whole. For that purpose, we used the deep spectrum obtained
when the SN was 309 days after maximum to extract spectra of the
host-galaxy light at different locations along the slit. We detected
narrow emission lines from \ion{H}{2} regions at different angular
distances from the SN. The extraction at the closest proximity to the
SN was done at a distance of $2{\farcs}2$, which means a
projected distance of 290 pc. We used the calibrations of
\citet{pettini04} to determine oxygen abundances based on flux ratios.
Adopting the $O3N2$ calibration, our measurements nearest the SN
resulted in $\log(\mathrm{O/H})+12=8.80$. If we adopt a solar
metallicity ($Z_\odot$) of $\log(\mathrm{O/H})+12=8.69 \pm 0.05$
\citep{asplund09}, the environment of SN~2010as has a metallicity of
$1.3\,Z_\odot$. The $N2$ index of \citet{pettini04} is not well-suited
for this SN because the 
measured line ratio is near the extreme of the relation. However, by
adopting both linear and cubic relations (their equations 1 and 2), we
obtained $\log(\mathrm{O/H})+12=8.75$ and $8.95$, which corresponds to
$1.1$ $Z_\odot$ and $1.8$ $Z_\odot$ respectively. This
suggests that the environment has super-solar metallicity. Near the SN
location, the measurement may be affected by the proximity of the
galaxy nucleus. We therefore performed the same measurement at longer
distances and found values of $\log(\mathrm{O/H})+12=8.74$ at
projected distances of up to 5 kpc from the nucleus. This provides
further evidence for the super-solar metallicity in the environment of
SN~2010as.

\subsection{Pre-Explosion Observations}
\label{sec:presn}

\noindent The field of SN~2010as was observed by the {\em HST} before
the explosion of the SN. Two images 
of NGC~6000 were retrieved from the Hubble Legacy Archive. The images
were obtained with the Wide Field 
Planetary Camera 2 (WFPC2) on 1996 February 19, using the F606W
filter (PI: Stiavelli) with a total exposure time of 600
seconds. Additional observations were found for a total exposure of
384 seconds in the F160W filter, obtained using the
Near Infrared Camera and Multi-Object Spectrometer (NICMOS) on 1997
July 5. We used these observations to 
study the nature of any possible source in the pre-explosion images
that could be associated with the progenitor of SN~2010as. 

In order to accurately determine the position of the SN in the WFPC2
images, we used high-resolution, adaptive optics observations 
obtained in the $K_s$ band using the NAOS-CONICA (NaCo) instrument
\citep{lenzen03,rousset03} at the Very Large Telescope of ESO by PI
Smartt on 2010 March 27, i.e., a few days before the SN reached
maximum light. Forty-eight dithered images of 60 seconds each were
processed and combined using the 
dedicated NaCo pipeline provided by ESO. We used IRAF routines in
order to locate common objects in the NaCo and WFPC2 F606W image and to
calculate the coordinate transformation. With five objects in common, 
we obtained a coordinate match with a precision of 16 mas, considering
the uncertainty in the plate solutions for both NaCo and WFPC2 images
(2 mas and 14 mas, respectively), and the geometric transformation
uncertainty of 6 mas, added in quadrature. The SN coordinates in the
NaCo image resulted as $\alpha = 15^{\mathrm{h}}49^{\mathrm{m}}49{\fs}228$, 
$\delta=-29^\circ23'09{\farcs}72$, in good agreement with previous
determinations. 

\begin{figure}[htpb]% Figure 13
\epsscale{1.0}
\plotone{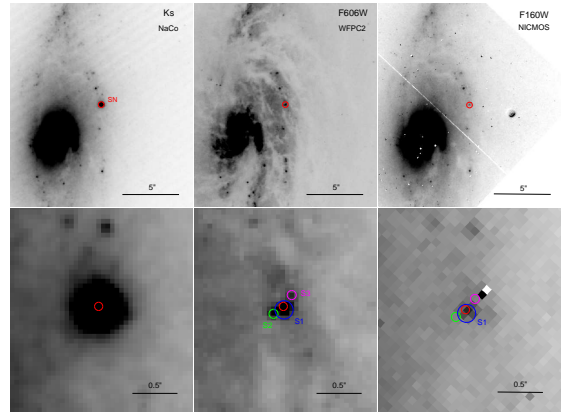}
\caption{Identification of the source at the SN location in
  pre-explosion {\em HST} images. North is up and East is to the left. 
   {\em Left column}: combined $K_s$-band 
  NaCo image from March 2010 showing the SN (red circle). {\em Middle
    column}: F606W-band WFPC2 image from February 1996. {\em Right
    column}: F160W-band NICMOS image from July 1997. In the zoomed-in
  images of the lower row
  the SN location is indicated by red circles with radius of three
  times the uncertainty in the relative astrometric solution. The
  locations of the sources S1, S2 and S3 detected in the F606W image
  are shown with blue, green and magenta circles, respectively. The
  sizes of the circles roughly indicate the source brightnesses. We
  associate S1 with the SN site. Only S1 is detected in the F160W
  image (right bottom panel).
  \label{fig:presn}}
\end{figure}

Using the above registration, we were able to identify a source that
was coincident with the SN location in both WFPC2 and NICMOS
images, as shown in Figure~\ref{fig:presn}. The WFPC2 image shows that
there were actually three objects in the vicinity of the SN location. The
brightest source (S1) is also the closest one to the SN location, at a
distance of $0{\farcs}04$---i.e., less than one pixel away, and less
than 5 pc away in projected distance. We
indicate the other two sources with S2 and S3 in order of decreasing F606W
brightness. S2 is located at $2.2$ pixels to the southeast of S1, that is at a
projected distance of $\approx$13 pc. S3 is located at $3.5$ pixels to
the NW of S1, with a projected distance of $\approx$21 pc. S1 appears
to have a slightly extended profile with FWHM of $1.9$ pixels---to be
compared with the point-source FWHM of $1.3$ pixels for this
instrument \citep{krist11}. In the F160W images only one source is
visible. This may be due to the 
coarser resolution of the NICMOS image (the PC chip of WFPC2 has a
plate scale of $0{\farcs}046$ per pixel, whereas the scale of NICMOS
is $0{\farcs}075$ per pixel). Indeed, the FWHM of the source is of 
$2.4$ pixels in the NICMOS image, and S2 and S3 would be located at $1.4$
and $2.2$ pixels from S1, respectively. We thus identify the source in the
NICMOS image with S1, although it may include flux from the unresolved
S2 and S3 sources. 

We performed PSF photometry on the F606W image using the IRAF DAOPHOT
package. The TinyTim PSF model failed to reproduce
the shape of S1 because it was slightly extended, so we needed to fit
the PSF using the source itself. We tested the photometry using
an aperture with a radius of 1 FWHM ($1.9$ pixels) and obtained
consistent results. Using the conversion formulae from
counts to magnitudes in the VEGAMAG system for the PC
chip\footnote{http://documents.stsci.edu/hst/wfpc2/documents/handbooks/dhb/wfpc2\_ch52.html},
we obtained, for S1, an apparent magnitude of
$m_{\mathrm{F606W}}({\mathrm S1})=22.566\pm0.052$ 
mag. For the other two sources, we obtained
$m_{\mathrm{F606W}}({\mathrm S2})=24.16\pm0.1$ mag and
$m_{\mathrm{F606W}}({\mathrm S3})=24.35\pm0.1$ mag. This means that S2 and S3 are
about four and five times fainter in F606W than S1, respectively. PSF
photometry was also performed with DAOPHOT on the F160W image 
using S1 to fit the PSF shape. With the conversion from count rate to
VEGAMAG system given by the NICMOS Handbook, we obtained
$m_{\mathrm{F160W}}({\mathrm S1})=21.297\pm0.455$ mag. 

We converted the F606W and F160W magnitudes of S1 to $V$ and $H$,
respectively. For this purpose, we first converted the VEGAMAG values
to AB magnitudes and then input them in the online conversion tool for
{\em HST}\footnote{http://www.stsci.edu/hst/nicmos/tools/conversion\_form.html}. 
For the conversion from AB magnitudes we adopted a blackbody spectrum
of 20000 K to model the source emission. Adopting temperatures between
10000 K and 30000 K to cover a range of possible progenitors, would
modify the result by at most $0.1$ mag in 
$V$, and $0.01$ mag in $H$. The resulting observed magnitudes are
$V({\mathrm S1})=22.49 \pm 0.11$ mag, and $H({\mathrm S1})= 21.22 \pm
0.46$ mag, considering the 
photometric and conversion uncertainties added in quadrature. The
observed color of S1 thus is $(V-H)=1.27 \pm 0.47$ mag.

Adopting a distance modulus of $32.16\pm 0.36$ mag for NGC~6000 (NED),
and correcting only for Galactic and host-galaxy extinction as derived
in Section~\ref{sec:col}, we obtained corrected absolute magnitudes
for S1 of $M^0_V({\mathrm S1}) = -10.77 \pm 0.41$ mag, and
$M^0_H({\mathrm S1}) = -11.08 \pm 0.58$ mag. 
The reddening-corrected color of S1 is therefore
$(V-H)^0=0.31 \pm 0.50$ mag. Although in principle the extinction
suffered by S1 may have been different from that of the SN,
the agreement between the reddening values obtained from the Balmer
decrement of the underlying \ion{H}{2} region and from the SN colors
suggests that both SN and its parent population were affected in the
same way by dust. If we consider the host-galaxy extinction law with
$R_V=3.1$ instead of the value of $R_V=1.5$ that was assumed for the
SN, this yields a corrected color for S1 of $(V-H)^0_{3.1}=0.08 \pm
0.57$ mag. Both values are consistent with each other within the
uncertainties. For consistency with the SN analysis, we adopted the
value for $R_V=1.5$ in the study of the underlying stellar population
that follows. 

\begin{figure}[htpb]% Figure 14
\epsscale{1.0}
\plotone{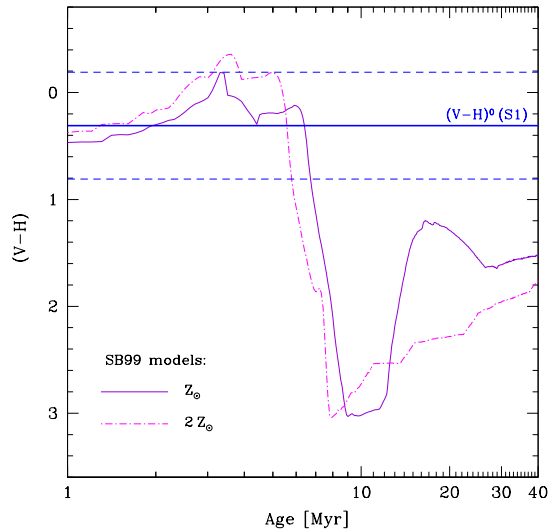}
\caption{Evolution of the $(V-H)$ color for synthetic stellar population
  models from Starburst99 assuming a single star-formation burst. The
  solid curve is for solar metallicity while the dot-dashed curve
  corresponds to twice-solar metallicity. Blue horizontal lines show the
  color of source S1 after correcting for reddening, with 1-$\sigma$
  uncertainties indicated by the dashed lines. Dashed horizontal
  lines indicate 1-$\sigma$ uncertainties in the observed color.
  \label{fig:sspvh}}
\end{figure}

We used the $(V-H)$ color and luminosity of source S1 to compare with
synthetic stellar population models in order to determine properties
of the progenitor cluster such as mass and age. For this purpose, we
adopted models from Starburst99 \citep{leitherer99} with solar
and twice-solar metallicity, encompassing the value 
measured near SN~2010as (see Section~\ref{sec:env}). In
Figure~\ref{fig:sspvh} we show the evolution of $(V-H)$ for an
instantaneous star-forming burst. From the reddening-corrected color
of S1, we obtained an age of the cluster of
$6.4^{+0.3}_{-3.0}$ Myr, considering the 1-$\sigma$ uncertainty in the
color. Assuming the progenitor of SN~2010as was the most massive star
in the cluster that produced source S1, we can derive its zero-age
main sequence (ZAMS) mass based on stellar evolution models. Adopting
single-star models of \citet{fagotto94a} for $Z_\odot$, the age range 
corresponds to masses of $M_{\mathrm{ZAMS}}>28.1$ $M_\odot$. Similar
results are obtained adopting evolutionary models  
by \citet{schaerer93}. Such masses are in the
regime of Wolf-Rayet (WR) stars. If we assume twice-solar metallicity, we
obtain an age of $5.6^{+0.2}_{-2.5}$, which implies a lower mass limit
of $M_{\mathrm{ZAMS}}>29.1$ $M_\odot$.

The synthetic population model with an age between 6 and 7 Myr,
solar metallicity, and a total luminosity of $M^0_V({\mathrm
  S1}) = -10.77$ mag would correspond to a total cluster mass of about
$2 \times 10^4$ $M_\odot$, containing $\sim$10 O-type
stars. A number of WR stars may still be present for such
cluster age.

We note that for the comparison with stellar population models described
in the previous paragraphs an instantaneous burst was assumed. If we
relax this assumption, we can obtain longer lifetimes and thus lower
masses. Assuming continuous star formation, we obtain upper limits for
the age of the cluster below $8.6$ Myr for solar metallicity, and below
$7.9$ Myr for $2\,Z_\odot$. These lifetimes translate into stellar masses of  
$M_{\mathrm{ZAMS}}>21.9$ $M_\odot$ and $M_{\mathrm{ZAMS}}>18.4$
$M_\odot$ for solar and twice-solar metallicity, respectively (note
that massive stars with high metallicity are shorter-lived than their
lower-metallicity counterparts, thus the lower-mass limit for the
$2\,Z_\odot$ case, even if the age of the cluster is younger). The
latter limits lie below the cutoff for WR stars. 

\subsection{Hydrodynamical models}
\label{sec:hydro}

\noindent One way to estimate the physical parameters of the SN
progenitor as well as the energy release and radioactive material
yield is by modelling the observed light curve and the photospheric
velocity ($v_{\mathrm{ph}}$) evolution. We used the bolometric light 
curve calculated in Section~\ref{sec:lbol} and the \ion{Fe}{2} line
velocity as a tracer for the photospheric velocity. Although in
general the latter is a good assumption for SE SNe \citep{branch02},
as we showed in Section~\ref{sec:specmax}, for SN~2010as the
photospheric velocity at 
early times may lie well below the \ion{Fe}{2} velocity (see also the
discussion below). For the hydrodynamical modelling we  
adopted an explosion time at the midpoint between the last
non-detection and the discovery time, $t_{\mathrm{exp}}$ (see
Section~\ref{sec:obs}).

We used the one-dimensional LTE radiation hydrodynamics code
previously presented by \citet{bersten11} to
calculate a set of SN models. Helium stars with different 
masses calculated from a single stellar evolutionary code
\citep{nomoto88} were adopted as pre-explosion configurations.
These initial models are devoid of hydrogen and 
have a compact structure with radius $R < 3\, R_\odot$. Note that
although the spectra of SN~2010as show evidence of a thin H envelope
(as shown in Section~\ref{sec:specmax}), the uncertainty on 
$t_{\mathrm{exp}}$ and the lack of an observed post-breakout cooling
phase in the light curves prevent us from
drawing conclusions about the extent of the envelope
\citep[see][]{bersten12}. We therefore focus our modelling only on the
$^{56}$Ni-powered part of the light curve where global 
properties such as explosion energy ($E$), ejected mass
($M_{\mathrm{ej}}$), and $^{56}$Ni mass can be estimated. 

\begin{figure}[htpb]% Figure 15A, Figure 15B
\begin{center}
\includegraphics[height=0.5\textwidth,angle=-90]{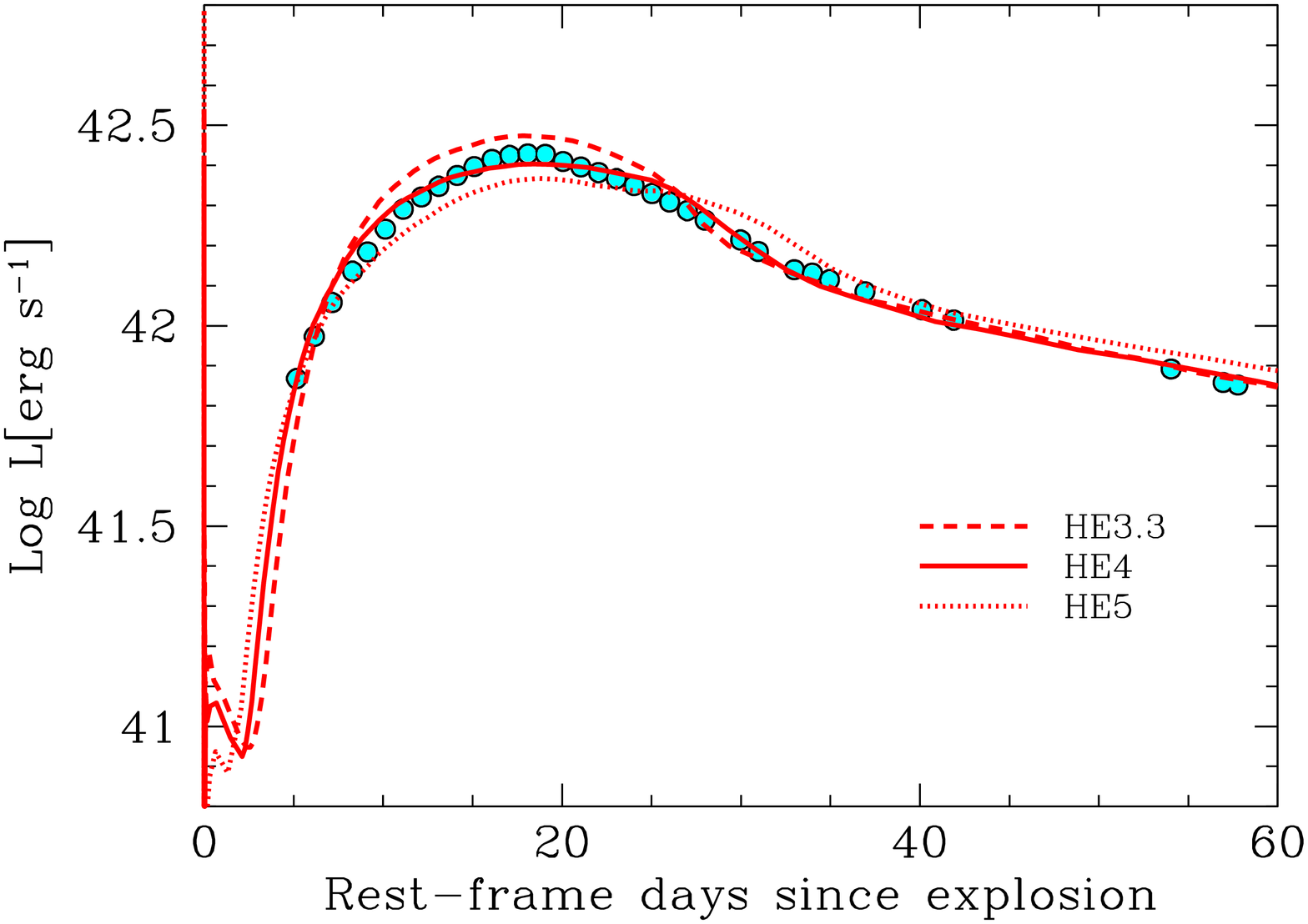}\\
\includegraphics[height=0.5\textwidth,angle=-90]{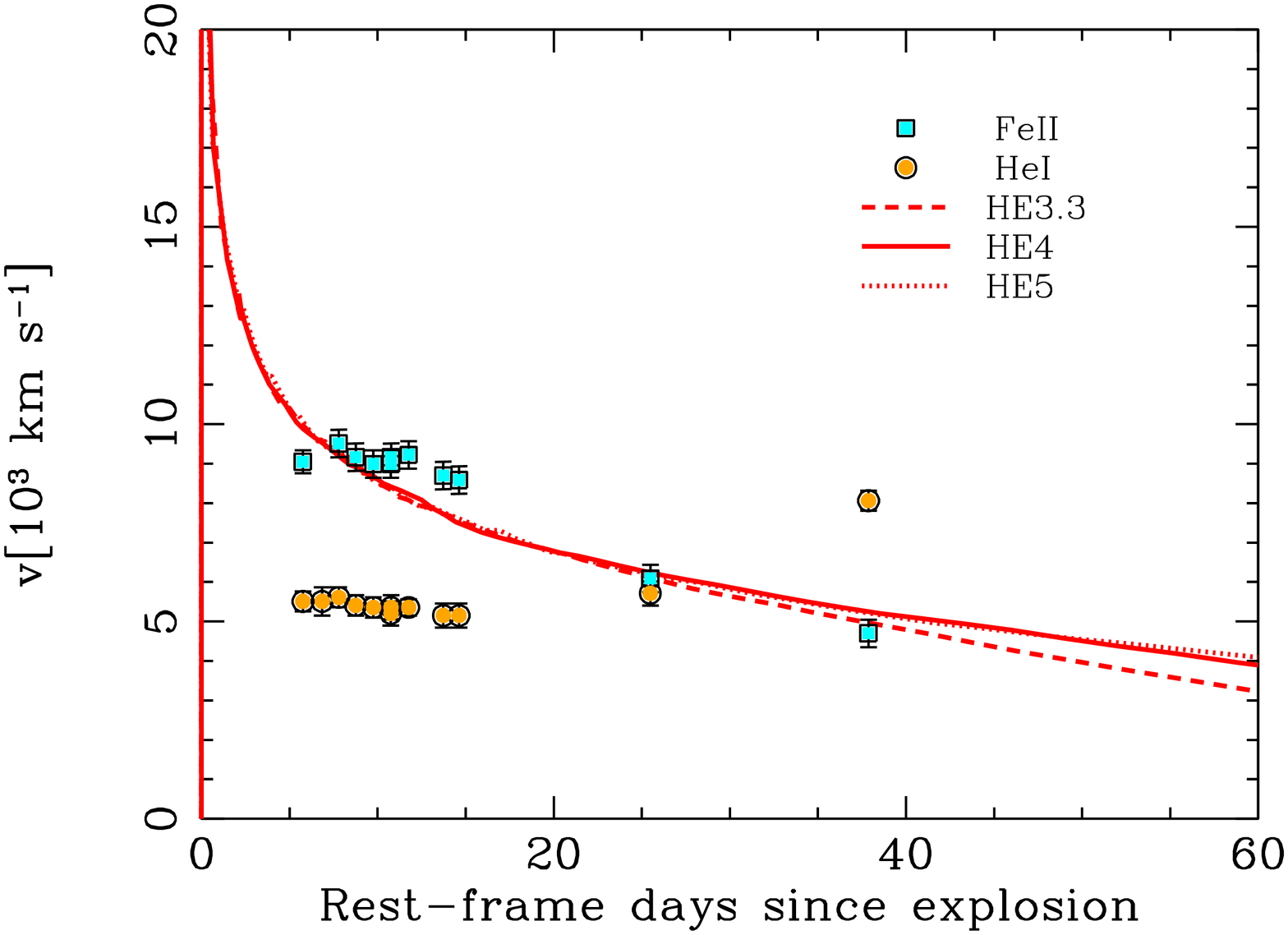}
\end{center}
\caption{{\em (Top panel)} Observed bolometric
  light curve of SN~2010as (circles) compared with the hydrodynamical
  models HE$3.3$ (dashed line), HE4
  (solid line) and HE5 (dotted line). {\em (Bottom panel)} Evolution
  of the photospheric velocity for the same models as above compared
  with measured line velocities of \ion{Fe}{2} (squares) and \ion{He}{1}
  (circles). \label{fig:hydro}}
\end{figure}

The sensitivity of the light curve and $v_{\mathrm{ph}}$ evolution to
$E$, $M_{\mathrm{ej}}$, $^{56}$Ni mass and $^{56}$Ni distribution has
been previously studied in 
\citet{bersten12}, and therefore it will not be described
here. Nevertheless, we note that, for a given progenitor mass,
$v_{\mathrm{ph}}$ imposes a stronger constraint on the energy than
does the light curve on the explosion energy. This is because of the weak
dependence of $v_{\mathrm{ph}}$ on the amount and distribution of
$^{56}$Ni. Figure~\ref{fig:hydro} shows 
the hydrodynamical modelling of SN~2010as for three different initial
configurations, namely for He stars of $3.3$ $M_\odot$ (HE$3.3$), 4
$M_\odot$ (HE4) and 5 $M_\odot$ (HE5). Such pre-SN configurations
correspond to single stars with main-sequence masses of 12
$M_\odot$, 15 $M_\odot$ and 18 $M_\odot$, respectively. From the
figure we see that the HE4 model provides the best
representation of the observations (light curve and velocity
together). The HE$3.3$ model is slightly overluminous, it produces an
acceptable approximation of the light-curve shape. A more massive
model (HE5) already produces a too wide light curve with an
underluminous peak. The failure of the massive
model resides in the difficulty of simultaneously fitting the light-curve
width and the low expansion velocities. In particular for SN~2010as,
even the \ion{Fe}{2} line velocities are relatively low at early
times. The width of the light curve grows with increasing pre-SN mass
and also with decreasing $E$. For massive models, the low observed
velocities require the energy to be relatively low, thus impeding the
possibility of reducing the light-curve width.

Here we adopt HE4 as our preferred model, and
the parameters used in the calculation shown in Figure~\ref{fig:hydro} are
an explosion energy of $E = 7 \times 10^{50}$ erg, an ejected mass of
$M_{\mathrm{ej}} = 2.5$ $M_\odot$\footnote{
   $M_{\mathrm{ej}} = M_{\mathrm{pre-SN}}-M_{\mathrm{cut}}$, where
   $M_{\mathrm{cut}}$ is the mass of the compact remnant assumed to be
   $1.5$ $M_\odot$}, and a $^{56}$Ni yield of about $0.12$
$M_\odot$. It is beyond the scope of this work to provide a
comprehensive analysis of the uncertainties in the physical
parameters. Given that the uncertainty in luminosity is nearly
systematic (see Section~\ref{sec:lbol}), the light curve may shift
vertically but its shape will remain fixed. The line velocities are
not affected by the distance uncertainty. Since $E$ and
$M_{\mathrm{ej}}$ both affect the shape of the light curve and the
velocities, we expect that most of
the luminosity uncertainty is transferred to the $^{56}$Ni mass. 
To reproduce the rise of the light curve we assumed $^{56}$Ni
mixing up to 95\% of the initial mass, although we note that this
value has a large uncertainty given the uncertainty on $t_{\mathrm{exp}}$.

It is interesting to note that even if we could reproduce the light
curve and the \ion{Fe}{2} velocity evolution reasonably well, none of our
models were able to reproduce the rather flat behavior of the
\ion{Fe}{2} velocity at early times, nor 
the even lower photospheric velocity derived from the SYNOW synthetic
spectra (see Section~\ref{sec:specmax}). That analysis
indicated that the photospheric velocity at early times could be
better represented by \ion{He}{1} or \ion{Mg}{2} lines than by
\ion{Fe}{2}. Assuming \ion{He}{1} velocity as the photospheric
velocity, a very weak explosion with $E \lesssim 2 
\times 10^{50}$ erg is required to reproduce the low $v_{\mathrm{ph}}$. However,
this leads to a very faint and wide light curve, in contradiction with the
observations, even for our lowest-mass progenitor. Therefore, we
conclude that if we use \ion{He}{1} 
velocities we cannot find a model that reproduces the light curve and
$v_{\mathrm{ph}}$ simultaneously under the current assumptions. In
this sense we emphasize that our calculations assume spherical
symmetry. It is possible that the peculiar behavior of \ion{He}{1}
velocities could be associated with departures from spherical symmetry
in the explosion \citep{tanaka09b}. 

We also analyzed the possibility that SN~2010as was powered by a 
magnetar instead of the typical $^{56}$Ni power. The motivations
to explore a magnetar as the energy source were, first, that SN~2010as showed
some spectroscopic similarities with SN~2005bf for which a magnetar 
was suggested to reproduce the unusual light-curve morphology
\citep{maeda07}, and, second, that magnetar models produce flat 
photospheric velocity evolution \citep[see][]{woosley10,kasen10,dessart12b}. We
included the magnetar energy in our hydrodynamical code following the
prescriptions of \citet{woosley10} and \citet{kasen10}. The magnetar  
model depends mainly on two parameters, the intensity of the magnetic
field and the initial rotation period of the neutron star. We tested 
different values of these parameters but we could not find a configuration
that could reproduce the light-curve
brightness and the rise time. In addition, we found that only powerful
magnetars can affect the dynamics of the SN ejecta, and thereby
produce a flat photospheric velocity evolution. However,
this type of models produces super-luminous supernovae which
are a few orders of magnitude more luminous than normal core-collapse
SNe. We therefore exclude the possibility of a magnetar as the main
source of power for SN~2010as. There may be other mechanisms besides a
magnetar that lead to the formation of a shell inside the ejecta, and
that may reproduce the observed velocity evolution. We plan to study
this interesting possibility in the future.

In conclusion, by approximating photospheric velocities with
\ion{Fe}{2} line velocities, our analysis suggested a progenitor star
with a He core of $\approx$4 $M_\odot$, which corresponds to a main
sequence mass of $\approx$15 $M_\odot$ (based on the adopted initial
model). The explosion released an energy of $E \approx 7 \times
10^{50}$ erg and produced $0.12$ $M_\odot$ of $^{56}$Ni. In addition, our
modelling disfavored models with He core mass above 5 $M_\odot$, which
implies a main sequence mass below 20 $M_\odot$. Such relatively low
masses suggest a binary progenitor for SN~2010as since stellar winds
alone would not produce enough mass loss to almost completely remove
the H-rich envelope \citep[see][]{puls08}. The formation of a dense
shell could be a solution for the flattening of the line velocity
evolution, although we did not provide a physical mechanism to account
for the existence of such dense shell. Departures from spherical
symmetry may also explain the low $v_{\mathrm{ph}}$.

\subsection{Progenitor Mass}
\label{sec:mass}

\noindent SN~2010as is one of the few H-poor SNe for which a
pre-explosion source could be identified. In this case, however, the
luminosity of the source is compatible with a massive stellar cluster rather
than the specific progenitor object (see Section~\ref{sec:presn}).
Therefore, the estimation of the progenitor mass was rather indirect
as it relied on assumptions on the star-formation history in the
cluster and on the model-dependent link between cluster age and
main-sequence mass of the most-massive star.
From the color of the source we derived an age of the cluster,
assuming a single star-formation burst. Provided the progenitor of
SN~2010as was the most massive star in the cluster, and assuming
single stellar evolution models, we were able to estimate the
main-sequence mass of the progenitor to be above 28--29 $M_\odot$,
depending on the adopted metallicity. This is
roughly compatible with the progenitor being a Wolf-Rayet star that
had lost most of its outer envelope via strong winds prior to
explosion. If we relax the assumption of a single stellar burst, the
age of the cluster could be older and that could imply masses slightly
below the range of WR stars \citep[see, e.g.,][]{langer12,groh13}.

On the other hand, the hydrodynamical modelling provides 
constraints on the progenitor mass at the moment of the explosion,
which can be associated to the ZAMS mass by assuming
some stellar evolution model. Our analysis of Section~\ref{sec:hydro}
favors a low-mass object with a pre-SN mass below 5 $M_\odot$,
in disagreement with a WR progenitor that would be more massive at
explosion. However, this analysis relies in part on the assumption that
the photospheric evolution can be traced by \ion{Fe}{2} line
velocities, which is questionable for SN~2010as. In addition, the
models failed to reproduce the observed flat velocity evolution. This 
may indicate that other mechanisms need to be considered or that some
of the model hypotheses, such as the spherical symmetry, may not be
valid. These difficulties make the conclusions from the hydrodynamic
modelling less accurate for this SN than what is usually the case.

The results from Sections~\ref{sec:presn} and \ref{sec:hydro} appear to be 
in contradiction. On the one hand, we found evidence for a young
progenitor age while, on the other hand, the exploding object appeared
to have a relatively low mass. Relaxing the main assumptions in each
case, i.e., the spherical symmetry in the hydrodynamical calculations,
and the single star-formation burst in the cluster age determination,
the disagreement can be improved. We further measured the flux ratio
of [\ion{Ca}{2}]~$\lambda$$\lambda$7291, to
[\ion{O}{1}]~$\lambda$$\lambda$6300, 6363, which is indicative of the
core mass, and thus of the ZAMS mass
\citep{fransson87,fransson89}. The spectrum at $+309$ days did not
cover the [\ion{O}{1}] completely. We assumed the profile to be
symmetric, and measure the line flux as twice the flux of the blue
component up to the central trough. The resulting
[\ion{Ca}{2}]/[\ion{O}{1}] ratio was 0.95. This value is similar to,
for example, those of SN~1996N and SN~2007Y \citep[][and references
  therein]{stritzinger09}, and it 
suggests a relatively low progenitor mass of $M_{\mathrm{ZAMS}}
\lesssim 20$ $M_\odot$.
Nevertheless, both pieces of evidence can be reconciled within the
scenario of a close binary progenitor system. The
  recent suggestion that about 70\% of the massive stars belong to
  interacting binary systems \citep{sana12} makes this a reasonable
  possibility. As shown in the calculations of \citet{yoon10},
interacting binaries with large 
initial masses of the primary stars ($\approx$25 $M_\odot$) may end
their evolution with relatively low pre-SN masses ($\approx$4
$M_\odot$) after losing most of the envelope via Roche-lobe
overflow. Moreover, the lifetime of the primary star would not be
significantly altered by the mass-transfer processes because it is
determined by the time-scale of nucleosynthesis in the stellar
core. Therefore, the age of the system at the moment of explosion
would also be compatible with the estimated cluster age.

\section{The Family of Flat-Velocity Type~IIb Supernovae}
\label{sec:family}

\subsection{Transitional Type~Ib/c SNe are Peculiar Type~IIb SNe}
\label{sec:spcomp}

\begin{figure}[htpb]% Figure 16
\epsscale{1.0}
\plotone{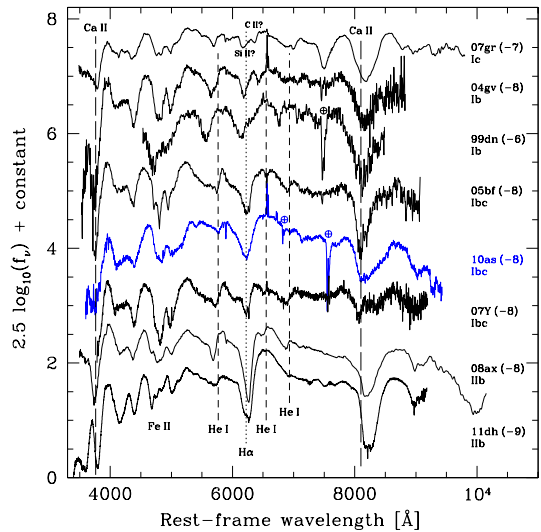}
\caption{Spectroscopic comparison between SN~2010as and other SE SNe
at six to nine days before maximum. Labels on the right-hand of each
spectrum indicate the SN name and type, and the epoch between
parentheses. Vertical lines show the location of absorption lines of
\ion{H}{1} ({\em dotted}), \ion{He}{1} ({\em short-dashed}) and
\ion{Ca}{2} ({\em long-dashed}), with a Doppler shift as measured for
SN~2010as. Earth symbols indicate partially or uncorrected telluric
absorptions. 
  \label{fig:spcompm008}}
\end{figure}

\begin{figure}[htpb]% Figure 17
\epsscale{1.0}
\plotone{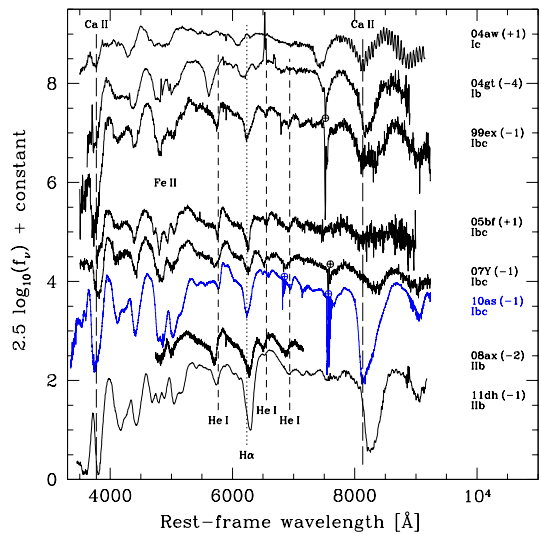}
\caption{Spectroscopic comparison between SN~2010as and other SE SNe
  around maximum light. Labels and symbols are as in
  Figure~\ref{fig:spcompm008}. 
  \label{fig:spcompm001}}
\end{figure}

\begin{figure}[htpb]% Figure 18
\epsscale{1.0}
\plotone{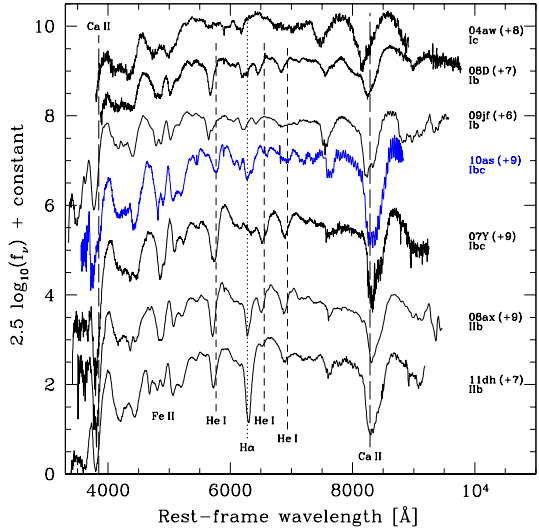}
\caption{Spectroscopic comparison between SN~2010as and other SE SNe
  at six to nine days after maximum light. Labels and symbols are as
  in Figure~\ref{fig:spcompm008}. 
  \label{fig:spcompp009}}
\end{figure}

\begin{figure}[htpb]% Figure 19
\epsscale{1.0}
\plotone{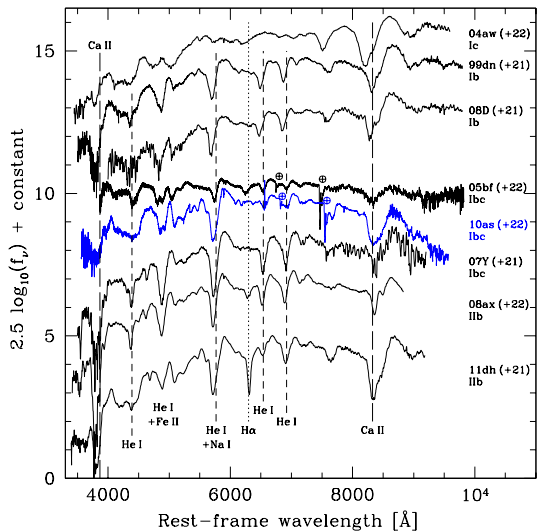}
\caption{Spectroscopic comparison between SN~2010as and other SE SNe
  at about three weeks after maximum light. Labels and symbols are as
  in Figure~\ref{fig:spcompm008}.
  \label{fig:spcompp022}}
\end{figure}

\begin{figure}[htpb]% Figure 20
\epsscale{1.0}
\plotone{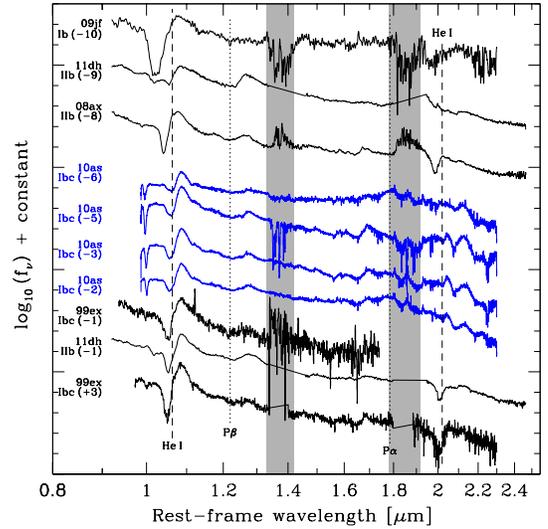}
\caption{Spectroscopic comparison between SN~2010as and other SE SNe
  in the NIR range. Labels on the left-hand side indicate the SN name,
  type and epoch relative to maximum light. Main \ion{He}{1} lines are
  indicated by vertical dashed lines located at a velocity of 5700 km
  s$^{-1}$. Also shown with dotted lines are the expected locations of
  P$\alpha$ and P$\beta$ at a velocity of 14500 km s$^{-1}$. The gray
  bands indicate the regions where atmospheric transmission drops 
  approximately below 50\%.
  \label{fig:spcompnir}}
\end{figure}

\noindent Figures~\ref{fig:spcompm008} to \ref{fig:spcompp022} show the
spectroscopic comparison in the 
optical between SN~2010as and other SE SNe at four different
epochs. Figure~\ref{fig:spcompnir}, additionally shows a comparison of
available pre-maximum spectra of SE SNe in the NIR range. Comparison
spectra were obtained from the {\em The Weizmann interactive supernova
  data repository} \citep{ofer12}\footnote{http://wiserep.weizmann.ac.il}.
The overall spectral evolution of SN~2010as makes this object most similar to
a group of transitional Type~Ib/c SNe that were first introduced with
the case of SN~1999ex \citep{hamuy02,stritzinger02}. Other objects
with similar spectroscopic characteristics are SN~2005bf
\citep{anupama05,tominaga05,folatelli06} and 2007Y \citep{stritzinger09}. The similarities
among these objects, in contrast with normal Type~Ib SNe, are most
clearly seen before maximum light, as shown in
Figures~\ref{fig:spcompm008} and \ref{fig:spcompm001}. The
distinguishing features among transitional Type~Ib/c SNe is the 
weakness of \ion{He}{1} lines, the strength of high-velocity \ion{Ca}{2}
absorptions, and the broad absorption near 6200 \AA\ that we have attributed
to H$\alpha$ for SN~2010as. Helium lines grow in strength until
maximum light, while high-velocity \ion{Ca}{2} components dominate
before maximum and disappear after. SN~2010as shows particularly
strong \ion{Ca}{2} absorptions at all times. As shown in
Figure~\ref{fig:spcompnir}, the pre-maximum NIR spectra of SN~2010as,
as well as those of SN~1999ex, show weaker \ion{He}{1} lines than the
comparison Type~Ib and IIb objects. 

After maximum light (Figure~\ref{fig:spcompp009}) helium lines are
clear in transitional Type~Ib/c SNe, while the absorption near 6200
\AA\ becomes weaker, as compared with typical Type~IIb objects, and it
appears blended with other lines. About three weeks after maximum
(Figure~\ref{fig:spcompp022}), SN~2010as 
and 2005bf remain similar while SN~2007Y shows well-developed
\ion{He}{1} lines. The spectroscopic similarity among the former two
objects is most striking if one considers the large differences in
light-curve properties and luminosities (see
Figure~\ref{fig:lbolcomp}). We note that at day 
$+22$ SN~2005bf\footnote{For SN~2005bf we consider the epochs relative
  to the first maximum in $B$ band.} is near the main
maximum in its light curve, while SN~2010as follows the usual
post-maximum decline of SE SNe.

The group of transitional Type~Ib/c SNe appears spectroscopically
closer to Type~IIb than to Type~Ib objects. Furthermore, the presence
of hydrogen seems to be ubiquitous among all these SNe. We have shown
in Section~\ref{sec:specmax} the identification of the
6200-\AA\ absorption with H$\alpha$ for SN~2010as. The same
identification has been suggested for
SN~1999ex \citep{branch06,elmhamdi06}, SN~2005bf
\citep{anupama05,folatelli06,parrent07}, and SN~2007Y
\citep{stritzinger09,maurer10}. The latter object may be a link between
transitional Type~Ib/c SNe and Type~IIb SNe, with lower H abundance,
as suggested by its resemblance with SN~2008ax (see
Figures~\ref{fig:spcompm008} through \ref{fig:spcompp022}).
Finally, H$\alpha$ is
present in the nebular spectra of SN~2005bf and SN~2007Y, and possibly in
SN~2010as (see Figure~\ref{fig:spcompneb}).
Hydrogen has also been suggested in early spectra of
Type~Ib SNe \citep[see][]{elmhamdi06}, although the confusion is
larger in this case between H$\alpha$ and \ion{Si}{2}~$\lambda$6355
\citep{hachinger12}. In the NIR range (see
Figure~\ref{fig:spcompnir}), although P$\alpha$ is difficult to
identify, hints of the presence of P$\beta$ are seen in SNe~2010as,
1999ex, 2011dh and 2008ax, but not in the Type~Ib SN~2009jf.

\begin{figure}[htpb]% Figure 21
\epsscale{1.0}
\plotone{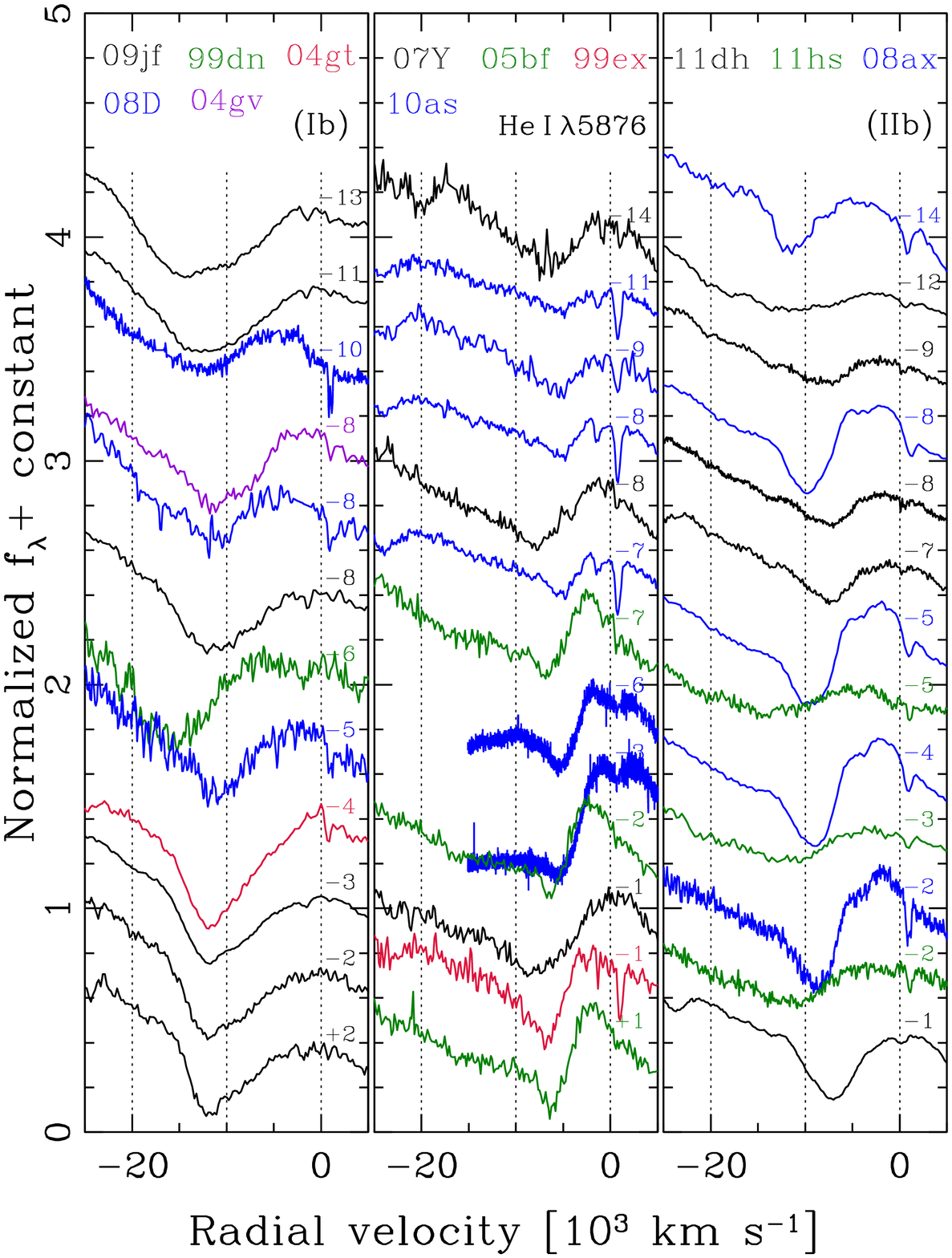}
\caption{Evolution of the \ion{He}{1}~$\lambda$5876 lines for SE
  SNe. Each graph shows 
  three columns, separated into Type~Ib, transitional Type~Ib/c, and
  Type~IIb SNe. The spectra are shown versus velocity relative to the rest
  wavelength of the line. Vertical dotted lines show 0, 10000 km
  s$^{-1}$ and 20000 km s$^{-1}$. Numerical labels on the right-hand
  side indicate the epoch of each spectrum in days with respect to
  $B$-band maximum light.
  \label{fig:helines}}
\end{figure}

The differences between transitional Type~Ib/c SNe and standard
Type~Ib objects are further
highlighted in Figures~\ref{fig:helines} and \ref{fig:halines}, where
we show the pre-maximum evolution of \ion{He}{1}~$\lambda$5876 and
H$\alpha$, respectively. The former line was chosen because it is the
strongest \ion{He}{1} line in the optical
range. \ion{He}{1}~$\lambda$5876 appears relatively 
weaker and at lower velocity in transitional Type~Ib/c and Type~IIb
SNe than in Type~Ib objects. The strength and shape of the H$\alpha$
profile also shows greater resemblance between the former two groups. 

From this analysis we conclude that transitional Type~Ib/c SNe
are in fact Type~IIb SNe. However, they can be distinguished from
standard Type~IIb objects. Their most evident peculiarity is the early
evolution of helium velocities. Figure~\ref{fig:velcomp} shows a
comparison of \ion{He}{1}~$\lambda$5876 velocities for a 
sample of SE SNe. All the transitional Type~Ib/c SNe can be
distinguished by their low and flat early-time helium velocity
evolution. This peculiar behavior was previously shown for SN~2005bf
by \citet{tominaga05}, and for SNe~1999ex and 2007Y by
\citet{taubenberger11}. Interestingly, the Type~IIb SN~2011dh shows the same 
peculiarity. The flat---sometimes increasing---behavior at early time is in
contrast with the typical velocity decline that is observed among SE
SNe, including other Type~IIb SNe and all Type~Ib objects \citep[see
  also][]{branch02}. A flattening of the \ion{He}{1}~$\lambda$5876
velocities can occur for SE SNe {\em after} maximum light, but this is
preceded by a steep decline. Moreover, the maximum-light 
\ion{He}{1}~$\lambda$5876 velocities are lower for the flat-velocity
objects than for the bulk of SE SNe. Values go from $\approx$5500 km
s$^{-1}$ for SN~2010as in the lower extreme, to $\approx$8000 km s$^{-1}$ for
SN~2007Y at the higher end. After maximum light,
\ion{He}{1}~$\lambda$5876 velocities for these objects tend to
increase. This may be due in part to contamination by \ion{Na}{1}. The
peculiarity of low and flat pre-maximum \ion{He}{1}~$\lambda$5876
velocity observed here cannot be explained by \ion{Na}{1}
contamination as this line is only about 800 km s$^{-1}$ to the red of
the helium feature. We
stress the fact that, for SN~2010as, the low and flat velocity
evolution was a general feature of the photosphere---as defined in
SYNOW---and not a peculiarity of the helium lines. 

This indicates the existence of a family of Type~IIb SNe that are
characterized by low and flat expansion velocities before maximum
light, most notably of helium lines, but presumably of the photosphere
as a whole. The members of such family are listed in
Table~\ref{tab:flatIIb}. We further propose to distinguish them from
standard Type~IIb objects with the name ``Type~fvIIb'' (standing for
``flat-velocity Type~IIb''). We find no
clear connection between this distinction based on early-time
velocities, and the separation between compact and extended Type~IIb
SNe proposed by \citet{chevalier10}, based on radio
observations. In terms of their expansion velocities, the prototypes
of compact (SN~2008ax) and extended (SN~1993J) Type~IIb SNe are very
similar. Given the post-maximum weakness of the H$\alpha$ line in
objects formerly identified as 
transitional Type~Ib/c SNe, this may indicate that they represent a
limit towards zero hydrogen abundance among Type~IIb SNe. The
remaining question is whether this tendency to zero hydrogen can be
physically associated with the peculiar behavior of helium lines or if
an additional mechanism should be invoked to account for such behavior.  

\subsection{On the Origin of Low and Flat Velocities}
\label{sec:nat}

\noindent The most striking feature of the family of
Type~fvIIb SNe is the persistence of low expansion
velocities. Considering also the presence of hydrogen and 
initial weakness of helium lines, we can characterize the spectroscopic
evolution of these objects as a transition from ``Type~IIc'' (i.e.,
with weak hydrogen and no helium) to Type~IIb. As noted by
\citet{dessart12a}, this very unique behavior is 
not easy to attain by the models, as it requires the 
presence of a small amount of hydrogen in the outer layers, and a
significant amount of helium that remains invisible at early times. In
such a picture, helium lines grow as the He-rich region becomes heated,
either by radioactive material, or by some other process that releases
energy in its vicinity. Such an alternative process may be the onset of a
magnetar \citep{dessart11}, although for SN~2010as and other
normal-luminosity SNe, this is unlikely (see Section~\ref{sec:hydro}).

If we consider radioactive material as the heat source, then the
degree of mixing with helium-rich material must play an important
role \citep{dessart12a}. The IIc--IIb transition requires that mixing
is moderate, so that helium excitation becomes significant only once the
photosphere reaches the $^{56}$Ni-rich layers. For Type~fvIIb SNe, in
order to maintain the velocity nearly 
constant, the increasing excitation of helium would possibly 
require an increasing $\gamma$-ray leakage from the
$^{56}$Ni-abundant region to the He-rich region, as suggested by
\citet{tominaga05} for SN~2005bf. From spectropolarimetry of that SN,
\citet{tanaka09b} further suggested that mixing of $^{56}$Ni with
helium occurred in an aspherical, clumpy manner.

\begin{figure}[htpb]% Figure 23
\epsscale{1.0}
\plotone{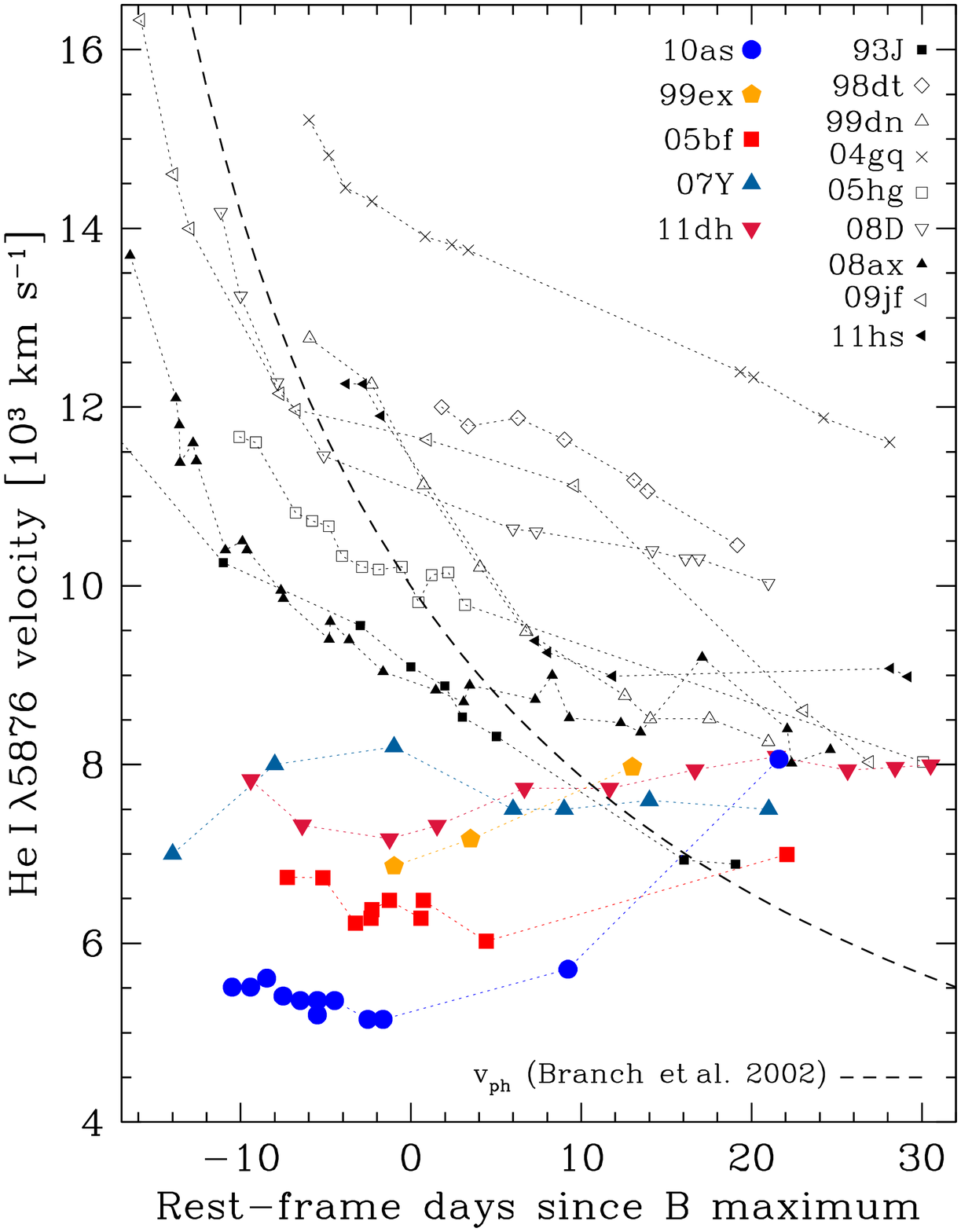}
\caption{Comparison of \ion{He}{1}~$\lambda$5876
  velocities for transitional Type~Ib/c SNe compared
  with other SE SNe. The black dashed
  curve shows the power-law evolution of $v_{\mathrm{ph}}$ that is
  typical of Type-Ib and Ic SNe, as derived by
  \citet{branch02}. Objects shown in color have flat helium velocity
  evolution before they cross the power-law $v_{\mathrm{ph}}$
  curve. Filled symbols indicate Type~IIb SNe while open symbols
  correspond to Type~Ib objects.
  \label{fig:velcomp}}
\end{figure}

The nearly constant photospheric velocity is suggestive of the
presence of a dense shell inside the SN ejecta. For different SNe, this
shell would be formed at different velocities, as shown in
Figure~\ref{fig:velcomp}. If the differences are intrinsic, they may
be related with a variation in the progenitor structure or a difference in the 
propagation of the explosive wave. The magnetar scenario, although it
can naturally produce a dense shell within the ejecta, is not favored
for SNe of normal luminosity and rise time. Alternatively, the
variation of velocity levels may not be an intrinsic property but a
consequence of a projection effect in an aspherical ejecta. In
Section~\ref{sec:specneb} we suggested an asymmetric distribution of
the inner O-rich material in SN~2010as, due to the double peak of the
[\ion{O}{1}] emission. However, this type of features
is not exclusively seen in SNe with flat velocities, and also the
geometry implied by the nebular spectrum may not coincide with 
that of the outer ejecta that is observed near maximum light. 

\begin{figure}[htpb]% Figure 22
\epsscale{1.0}
\plotone{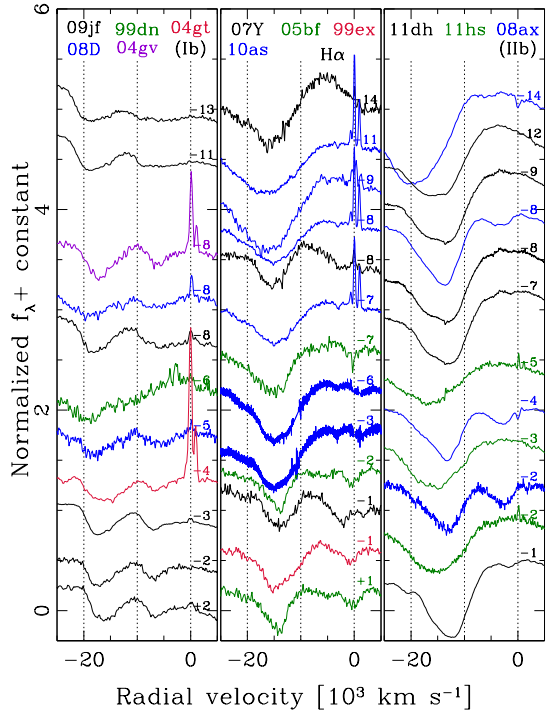}
\caption{Evolution of the H$\alpha$ lines for SE SNe. Each graph shows
  three columns, separated into Type~Ib, transitional Type~Ib/c, and
  Type~IIb SNe. The spectra are shown versus velocity relative to the rest
  wavelength of the line. Vertical dotted lines show 0, 10000 km
  s$^{-1}$ and 20000 km s$^{-1}$. Numerical labels on the right-hand
  side indicate the epoch of each spectrum in days with respect to
  $B$-band maximum light.
  \label{fig:halines}}
\end{figure}

The helium velocity, and presumably the photospheric velocity,
is different for different flat-velocity objects, as shown in
Figure~\ref{fig:velcomp}. The following sequence of increasing
helium---and presumably photospheric---velocity 
is observed: SN~2010as--2005bf--1999ex--2011dh--2007Y. There is a hint of
decreasing luminosities in the same direction, as seen in
Table~\ref{tab:flatIIb} and Figure~\ref{fig:lbolcomp}. A larger sample
is required to investigate whether such sequence becomes a correlation
between luminosity and helium velocity. The large spectroscopic sample
recently released by \citet{modjaz14} may serve to this
purpose. Similarly, an increased sample 
and spectral modelling may provide clues as to whether hydrogen 
abundances reproduce the same sequence. From the strength of
H$\alpha$ in the photospheric phase, this is not clear (see
Figure~\ref{fig:halines}), however the observed strength of H$\alpha$ in
the nebular phase suggests that the amount hydrogen seen at late-time
grows in the expected order.

\section{CONCLUSIONS}
\label{sec:concl}

\noindent We have used extensive observations of the SE SN~2010as to
characterize its photometric and spectroscopic evolution. The
long-wavelength base of the photometry, from $B$ band to $K$ band,
allowed us to compute several color indices and, from a comparison
with colors of a sample of SE SNe, determine a non-standard reddening
characterized by a low total-to-selective extinction coefficient of
$R_V=1.5$. The optical and NIR photometry was used to integrate a
bolometric light curve that was used to compare with hydrodynamic
modelling. The spectra, obtained with a nearly nightly cadence before
maximum light and with NIR coverage, allowed the identification of the
main ions in the SN ejecta, and their approximate distribution in
velocity space. Strong evidence of the presence of hydrogen was
found. The evolution of the photospheric velocity was found to be
peculiarly low and nearly constant in time since the earliest
spectroscopy epochs at $-11$ days with respect to maximum light.

SN~2010as was associated with a young, massive stellar cluster present
in archival {\em HST} images. This suggested a large ZAMS mass for the
progenitor, roughly compatible with a WR star, under the assumption of
a single star burst. In contrast with that, our hydrodynamical
modelling showed that the explosion energy and 
pre-SN mass should be relatively low. Our best match model has 
$E \approx 7 \times 10^{50}$ erg, $M_{\mathrm{pre-SN}} \approx 4$
$M_\odot$, and a $^{56}$Ni yield of $\approx$$0.12$
$M_\odot$. However, the model fails to reproduce the observed flat and
low velocity evolution. This may be due
to a limitation of our one-dimensional prescription, or to a missing 
physical mechanism that may form a dense shell in the SN ejecta. 
In any case, the seeming inconsistency can be readily solved by
considering an interacting binary progenitor scenario, where an
initially massive star looses its envelope through mass transfer to
the secondary star and reaches the pre-supernova stage at a young age
but with a relatively low mass \citep{yoon10,bersten14}. 

SN~2010as belongs to a small family of SE SNe listed in
Table~\ref{tab:flatIIb} that show peculiar spectroscopic properties,
such as the transition from almost He-free to He-rich spectra,
accompanied by low and nearly constant photospheric 
velocities near maximum light. Although some of these objects were
previously classified as ``transitional Type~Ib/c'' SNe, we have shown
compelling evidence of the ubiquitous presence of hydrogen in this
group, which calls for a Type~IIb classification. However, they show a
peculiar velocity evolution, so we have distinguished them as
``flat-velocity Type~IIb'' (Type~fvIIb) SNe. This group might
represent the limit to zero hydrogen among Type~IIb SNe, although this
needs to be verified with spectral modelling. 

The distinguishing property of approximately constant velocity for this
family of SNe is suggestive of a dense shell in the ejecta. Velocities
of different SNe are different and span the range between 6000 and
8000 km s$^{-1}$, with SN~2010as showing the lowest velocity in the
group. The luminosities of Type~fvIIb objects show a wide variety
that might be correlated with the expansion velocities. This leaves
the open questions of the origin of the low and flat velocities, and
of its possible connection, or lack of, with luminosity. Continued
studies of flat-velocity SNe may help to understand the mechanism
behind their observed peculiarities, whether of physical origin
related with the progenitor or the explosion properties, or as a
consequence of a geometrical configuration. 

\acknowledgments 
This research is supported by the World Premier International Research Center
Initiative (WPI Initiative), MEXT, Japan. 
G.~F.\ is thankful to Masaomi Tanaka and Sergei Blinnikov for fruitful
discussion. 
We thank members of the GROND team: Marco Nardini, Robert Filgas,
Patricia Schady, Paulo Afonso, Ana Nicuesa Guelbenzu, and Sebastian
Schmidl. 
G.~F.\ acknowledges financial support by Grant-in-Aid for Scientific
Research for Young Scientists (23740175). 
G.~P.\ acknowledges support by the Proyecto FONDECYT 11090421.
M.~H., G.~P., F.~F., and F.~O~.E.\ acknowledge support provided by the
Millennium Institute of Astrophysics (MAS) through grant IC120009 of
the Programa Iniciativa Cientifica Milenio del Ministerio de Economia,
Fomento y Turismo de Chile. 
F.~F.\ further acknowledges support from FONDECYT through start-up
grant 11130228, H.~K.\ through postdoctoral grant 3140563,
H.~K.\ through postdoctoral grant 3140563, and F.~O.~E.\ through
postdoctoral grant 3140326. 
M.~D.~S.\ and C.~C.\ acknowledge support provided by the Danish Agency
for Science and Technology and Innovation realized through a Sapere
Aude Level 2 grant. 
M.~D.~S.\ and S.~H.\ acknowledge funding provided by the Instrument
Center for Danish Astrophysics (IDA). 
K.~N.\ is supported by the Grant-in-Aid for Scientific Research
(23224004, 23540262, 26400222). 
K.~M.\ is supported by the Grant-in-Aid for Scientific Research
(23740141, 26800100).
Part of the GROND funding (both hardware and personnel) was generously
granted from the Leibniz-Prize to Prof.~G.\ Hasinger, Deut\-sche
For\-schungs\-ge\-mein\-schaft (DFG) grant HA 1850/28--1. 
This research has made use of the NASA/IPAC Extragalactic Database (NED)
which is operated by the Jet Propulsion Laboratory, California
Institute of Technology, under contract with the National
Aeronautics and Space Administration.

%\appendix
%\input{appendix.tex}

\clearpage
\begin{deluxetable}{lcccccc} 
\tabletypesize{\small} 
\tablecolumns{7} 
\tablewidth{0pt} 
\tablecaption{Landolt-system $BVRI$ photometric sequence of field stars around SN~2010as.\label{tab:lseq}} 
\tablehead{ 
\colhead{Star} & \colhead{RA} & \colhead{Dec.} & \colhead{$B$} & \colhead{$V$} & \colhead{$R$} & \colhead{$I$} \\ 
\colhead{ID} & \colhead{(h:m:s)} & \colhead{($^\circ$:$'$:$''$)} & \colhead{(mag)} & \colhead{(mag)} & \colhead{(mag)} & \colhead{(mag)} 
} 
\startdata
1  & 15:49:47.411 & $-29$:22:00.21 & $15.303(035)$ & $14.723(012)$ & $14.638(657)$ & $14.004(034)$ \\ 
2  & 15:49:50.596 & $-29$:21:30.22 & $16.762(119)$ & $16.091(012)$ & $15.615(044)$ & $15.191(035)$ \\ 
3  & 15:49:59.370 & $-29$:22:58.17 & $14.393(053)$ & $13.195(026)$ & $12.541(028)$ & $11.968(015)$ \\ 
4  & 15:49:59.634 & $-29$:22:46.47 & $17.030(057)$ & $16.223(022)$ & $15.713(026)$ & $15.220(017)$ \\ 
5  & 15:50:01.984 & $-29$:23:02.57 & $15.087(061)$ & $13.764(024)$ & $13.024(023)$ & $12.345(013)$ \\ 
6  & 15:49:53.249 & $-29$:24:32.93 & $15.996(102)$ & $14.835(047)$ & $14.126(061)$ & $13.451(043)$ \\ 
7  & 15:49:43.791 & $-29$:24:39.59 & $16.004(062)$ & $15.285(020)$ & $14.843(024)$ & $14.459(025)$ \\ 
8  & 15:49:44.342 & $-29$:23:52.40 & $16.950(043)$ & $16.258(020)$ & $15.811(030)$ & $15.408(021)$ \\ 
9  & 15:49:37.469 & $-29$:24:01.19 & $15.229(054)$ & $14.604(017)$ & $14.211(029)$ & $13.845(021)$ \\ 
10 & 15:49:52.330 & $-29$:20:26.20 & $15.280(010)$ & $16.169(015)$ & $15.675(054)$ & $15.534(528)$ \\ 
11 & 15:49:53.861 & $-29$:20:42.09 & $15.280(010)$ & $17.566(011)$ & $16.868(057)$ & $16.332(027)$ \\ 
12 & 15:49:56.994 & $-29$:26:00.37 & $14.646(072)$ & $13.889(019)$ & $13.443(031)$ & $13.061(018)$ \\ 
13 & 15:50:05.146 & $-29$:25:45.93 & $16.380(043)$ & $14.993(018)$ & $14.215(017)$ & $13.582(010)$ \\ 
14 & 15:49:47.942 & $-29$:24:53.56 & $17.487(085)$ & $16.669(027)$ & $16.170(028)$ & $15.727(027)$ \\ 
15 & 15:49:46.103 & $-29$:25:10.48 & $17.517(076)$ & $16.857(024)$ & $16.424(030)$ & $16.028(030)$ \\ 
16 & 15:49:39.244 & $-29$:24:13.54 & $17.657(065)$ & $16.794(024)$ & $16.251(029)$ & $15.747(023)$ \\ 
17 & 15:49:37.704 & $-29$:21:58.24 & $15.431(057)$ & $14.824(010)$ & $14.399(039)$ & $14.009(027)$ \\ 
18 & 15:49:39.608 & $-29$:20:27.26 & $15.280(010)$ & $15.982(059)$ & $15.756(497)$ & $15.089(040)$ \\ 
19 & 15:49:37.335 & $-29$:20:44.98 & $15.280(010)$ & $16.785(012)$ & $16.305(050)$ & $15.885(037)$ \\ 
\enddata 
\tablecomments{Uncertainties in parentheses are given in thousandth of a magnitude.}
\end{deluxetable} 
 % Table 1

\clearpage
\begin{deluxetable}{lcccccc} 
\tabletypesize{\small} 
\tablecolumns{7} 
\tablewidth{0pt} 
\tablecaption{Sloan-system $g'r'i'z'$ photometric sequence of field stars around SN~2010as.\label{tab:sseq}} 
\tablehead{ 
\colhead{Star} & \colhead{RA} & \colhead{Dec.} & \colhead{$g'$} & \colhead{$r'$} & \colhead{$i'$} & \colhead{$z'$} \\ 
\colhead{ID} & \colhead{(h:m:s)} & \colhead{($^\circ$:$'$:$''$)} & \colhead{(mag)} & \colhead{(mag)} & \colhead{(mag)} & \colhead{(mag)} 
} 
\startdata
1  & 15:49:47.411 & $-29$:22:00.21 & $14.988(027)$ & $14.584(013)$ & $14.419(025)$ & $14.374(031)$ \\ 
2  & 15:49:50.596 & $-29$:21:30.22 & $16.450(028)$ & $15.873(014)$ & $15.628(025)$ & $15.545(030)$ \\ 
3  & 15:49:59.370 & $-29$:22:58.17 & $13.735(030)$ & $12.836(014)$ & $12.457(032)$ & $12.264(035)$ \\ 
4  & 15:49:59.634 & $-29$:22:46.47 & $16.617(033)$ & $15.978(013)$ & $15.669(027)$ & $15.540(047)$ \\ 
5  & 15:50:01.984 & $-29$:23:02.57 & $14.362(033)$ & $13.342(012)$ & $12.852(029)$ & $12.599(037)$ \\ 
6  & 15:49:53.249 & $-29$:24:32.93 & \nodata & \nodata & \nodata & \nodata \\ 
7  & 15:49:43.791 & $-29$:24:39.59 & $15.616(026)$ & $15.089(012)$ & $14.881(032)$ & $14.813(035)$ \\ 
8  & 15:49:44.342 & $-29$:23:52.40 & $16.591(027)$ & $16.063(022)$ & $15.836(051)$ & $15.744(059)$ \\ 
9  & 15:49:37.469 & $-29$:24:01.19 & $14.902(025)$ & $14.445(015)$ & $14.262(025)$ & $14.210(028)$ \\ 
10 & 15:49:52.330 & $-29$:20:26.20 & $16.563(034)$ & $15.929(015)$ & $15.692(031)$ & $15.612(027)$ \\ 
11 & 15:49:53.861 & $-29$:20:42.09 & $18.121(040)$ & $17.166(021)$ & $16.792(033)$ & $16.618(039)$ \\ 
12 & 15:49:56.994 & $-29$:26:00.37 & $14.216(030)$ & $13.692(013)$ & $13.483(032)$ & $13.424(031)$ \\ 
13 & 15:50:05.146 & $-29$:25:45.93 & $15.637(028)$ & $14.546(013)$ & $14.091(037)$ & $13.844(043)$ \\ 
14 & 15:49:47.942 & $-29$:24:53.56 & $17.062(030)$ & $16.423(012)$ & $16.165(018)$ & $16.073(031)$ \\ 
15 & 15:49:46.103 & $-29$:25:10.48 & $17.176(032)$ & $16.665(014)$ & $16.445(021)$ & $16.364(030)$ \\ 
16 & 15:49:39.244 & $-29$:24:13.54 & $17.232(040)$ & $16.516(015)$ & $16.198(021)$ & $16.065(029)$ \\ 
17 & 15:49:37.704 & $-29$:21:58.24 & $15.133(029)$ & $14.649(014)$ & $14.432(026)$ & $14.360(028)$ \\ 
18 & 15:49:39.608 & $-29$:20:27.26 & $16.287(028)$ & $15.745(013)$ & $15.513(032)$ & $15.421(031)$ \\ 
19 & 15:49:37.335 & $-29$:20:44.98 & $17.170(037)$ & $16.572(011)$ & $16.315(027)$ & $16.230(035)$ \\ 
\enddata 
\tablecomments{Uncertainties in parentheses are given in thousandth of a magnitude.}
\end{deluxetable} 
 % Table 2

\clearpage
\begin{deluxetable}{lccccc} 
\tabletypesize{\small} 
\tablecolumns{6} 
\tablewidth{0pt} 
\tablecaption{Landolt-system $BVRI$ magnitudes of SN~2010as from PROMPT.\label{tab:lsn}} 
\tablehead{ 
\colhead{UT date} & \colhead{JD} & \colhead{$B$} & \colhead{$V$} & \colhead{$R$} & \colhead{$I$} \\ 
\colhead{(yyyy-mm-dd)} & \colhead{($-2400000$)} & \colhead{(mag)} & \colhead{(mag)} & \colhead{(mag)} & \colhead{(mag)} 
} 
\startdata
2010-03-20.20 & 55275.70 & $18.166(066)$ & $16.948(029)$ & $16.560(025)$ & $16.294(030)$ \\ 
2010-03-21.20 & 55276.70 & $17.730(057)$ & $16.779(026)$ & $16.330(022)$ & $16.007(026)$ \\ 
2010-03-22.20 & 55277.70 & $17.523(046)$ & $16.564(024)$ & $16.113(020)$ & $15.736(023)$ \\ 
2010-03-24.10 & 55279.60 & $17.210(043)$ & $16.239(024)$ & $15.790(019)$ & $15.550(020)$ \\ 
2010-03-25.20 & 55280.70 & $16.995(039)$ & $16.118(022)$ & $15.719(018)$ & $15.377(018)$ \\ 
2010-03-25.20 & 55280.70 & $16.974(038)$ & $16.130(019)$ & $15.703(017)$ & $15.372(020)$ \\ 
2010-03-26.30 & 55281.80 & $16.859(035)$ & $15.991(021)$ & $15.545(017)$ & $15.314(018)$ \\ 
2010-03-28.20 & 55283.70 & $16.704(032)$ & $15.856(017)$ & $15.445(014)$ & $15.137(017)$ \\ 
2010-03-28.20 & 55283.70 & $16.722(033)$ & $15.858(020)$ & $15.440(016)$ & $15.158(016)$ \\ 
2010-03-29.20 & 55284.70 & $16.634(035)$ & \nodata & \nodata & \nodata \\ 
2010-03-30.20 & 55285.70 & $16.605(030)$ & $15.744(016)$ & $15.308(015)$ & $15.021(016)$ \\ 
2010-03-30.20 & 55285.70 & $16.632(030)$ & $15.719(020)$ & $15.303(016)$ & $15.033(016)$ \\ 
2010-03-31.20 & 55286.70 & $16.580(031)$ & $15.709(016)$ & $15.253(014)$ & $14.964(016)$ \\ 
2010-04-01.20 & 55287.70 & $16.580(031)$ & $15.678(016)$ & $15.183(014)$ & $14.895(016)$ \\ 
2010-04-02.20 & 55288.70 & $16.593(035)$ & \nodata & \nodata & \nodata \\ 
2010-04-03.20 & 55289.70 & $16.656(032)$ & $15.646(016)$ & $15.130(013)$ & \nodata \\ 
2010-04-04.20 & 55290.70 & $16.815(036)$ & $15.680(016)$ & $15.153(013)$ & $14.849(015)$ \\ 
2010-04-06.20 & 55292.70 & $17.013(039)$ & $15.761(016)$ & $15.182(013)$ & $14.824(015)$ \\ 
2010-04-08.20 & 55294.70 & $17.250(043)$ & $15.859(019)$ & $15.189(015)$ & $14.829(014)$ \\ 
2010-04-09.20 & 55295.70 & $17.351(044)$ & \nodata & \nodata & \nodata \\ 
2010-04-10.20 & 55296.70 & $17.469(045)$ & $15.989(020)$ & $15.329(015)$ & $14.918(015)$ \\ 
2010-04-12.30 & 55298.80 & $17.647(054)$ & \nodata & \nodata & \nodata \\ 
2010-04-15.20 & 55301.70 & $17.940(059)$ & $16.409(022)$ & $15.620(016)$ & $15.103(017)$ \\ 
2010-04-19.20 & 55305.70 & $18.289(067)$ & $16.609(025)$ & $15.804(018)$ & $15.195(018)$ \\ 
2010-04-21.20 & 55307.70 & $18.405(067)$ & $16.775(027)$ & $15.922(019)$ & $15.276(019)$ \\ 
2010-04-26.20 & 55312.70 & $18.724(090)$ & $16.870(028)$ & $16.149(021)$ & $15.396(020)$ \\ 
2010-05-08.30 & 55324.80 & $18.950(084)$ & \nodata & \nodata & \nodata \\ 
2010-05-12.10 & 55328.60 & $18.992(102)$ & $17.385(037)$ & $16.726(028)$ & $15.972(028)$ \\ 
2010-05-19.10 & 55335.60 & $19.000(087)$ & \nodata & \nodata & \nodata \\ 
2010-05-30.10 & 55346.60 & \nodata & $17.710(044)$ & $17.226(036)$ & $16.450(033)$ \\ 
2010-06-05.10 & 55352.60 & $19.194(096)$ & \nodata & \nodata & \nodata \\ 
2010-06-12.00 & 55359.50 & $19.345(118)$ & \nodata & \nodata & \nodata \\ 
2010-07-12.00 & 55389.50 & \nodata & $18.412(060)$ & $18.189(059)$ & $17.199(044)$ \\ 
\enddata 
\tablecomments{Uncertainties in parentheses are given in thousandth of a magnitude.}
\end{deluxetable} 
 % Table 3

\clearpage
\begin{deluxetable}{lccccc} 
\tabletypesize{\small} 
\tablecolumns{6} 
\tablewidth{0pt} 
\tablecaption{Sloan-system $g'r'i'z'$ magnitudes of SN~2010as from PROMPT.\label{tab:ssn}} 
\tablehead{ 
\colhead{UT date} & \colhead{JD} & \colhead{$g'$} & \colhead{$r'$} & \colhead{$i'$} & \colhead{$z'$} \\ 
\colhead{(yyyy-mm-dd)} & \colhead{($-2400000$)} & \colhead{(mag)} & \colhead{(mag)} & \colhead{(mag)} & \colhead{(mag)} 
} 
\startdata
2010-03-19.73 & 55275.24 & $17.544(034)$ & \nodata & \nodata & \nodata \\ 
2010-03-20.91 & 55276.40 & $17.316(035)$ & $16.493(021)$ & \nodata & \nodata \\ 
2010-03-21.75 & 55277.25 & $17.039(028)$ & \nodata & \nodata & \nodata \\ 
2010-03-23.66 & 55279.15 & $16.689(022)$ & $16.026(016)$ & $15.979(019)$ & $16.115(032)$ \\ 
2010-03-24.74 & 55280.24 & $16.546(020)$ & $15.920(015)$ & $15.863(017)$ & $15.970(029)$ \\ 
2010-03-25.79 & 55281.29 & $16.426(020)$ & $15.796(015)$ & $15.758(016)$ & $15.817(027)$ \\ 
2010-03-26.73 & 55282.22 & $16.362(018)$ & $15.725(014)$ & $15.689(016)$ & $15.719(026)$ \\ 
2010-03-27.73 & 55283.24 & $16.271(019)$ & $15.641(013)$ & $15.630(015)$ & $15.651(025)$ \\ 
2010-03-28.74 & 55284.24 & $16.233(018)$ & $15.582(013)$ & $15.511(015)$ & $15.532(025)$ \\ 
2010-03-29.75 & 55285.25 & $16.179(017)$ & $15.516(014)$ & $15.486(016)$ & $15.497(024)$ \\ 
2010-03-30.73 & 55286.23 & $16.151(017)$ & $15.461(013)$ & $15.387(016)$ & $15.437(025)$ \\ 
2010-03-31.73 & 55287.22 & $16.140(017)$ & $15.448(013)$ & $15.408(015)$ & $15.390(025)$ \\ 
2010-04-01.72 & 55288.22 & $16.145(018)$ & $15.384(020)$ & $15.346(026)$ & $15.357(051)$ \\ 
2010-04-02.72 & 55289.22 & $16.146(018)$ & $15.400(013)$ & $15.335(018)$ & $15.374(028)$ \\ 
2010-04-03.71 & 55290.21 & $16.203(017)$ & $15.381(012)$ & $15.276(013)$ & $15.321(022)$ \\ 
2010-04-04.71 & 55291.21 & \nodata & $15.394(012)$ & $15.265(014)$ & $15.345(024)$ \\ 
2010-04-05.71 & 55292.21 & $16.327(019)$ & $15.442(013)$ & $15.300(013)$ & $15.311(021)$ \\ 
2010-04-06.71 & 55293.21 & $16.371(020)$ & $15.457(012)$ & $15.277(014)$ & $15.354(022)$ \\ 
2010-04-07.73 & 55294.23 & $16.471(024)$ & \nodata & \nodata & \nodata \\ 
2010-04-08.71 & 55295.21 & $16.587(022)$ & $15.517(012)$ & $15.353(014)$ & $15.354(024)$ \\ 
2010-04-10.70 & 55297.20 & $16.762(022)$ & $15.649(015)$ & \nodata & \nodata \\ 
2010-04-11.70 & 55298.20 & $16.825(024)$ & $15.702(013)$ & $15.446(015)$ & $15.481(023)$ \\ 
2010-04-14.69 & 55301.19 & $17.113(026)$ & $15.881(015)$ & $15.594(015)$ & $15.615(025)$ \\ 
2010-04-18.67 & 55305.17 & $17.334(028)$ & $16.101(016)$ & $15.689(016)$ & $15.848(027)$ \\ 
2010-04-20.67 & 55307.17 & $17.512(031)$ & $16.238(017)$ & $15.796(017)$ & $15.836(028)$ \\ 
2010-04-25.66 & 55312.16 & $17.758(036)$ & $16.433(021)$ & $16.047(020)$ & $15.988(031)$ \\ 
2010-05-04.62 & 55321.12 & \nodata & $16.852(025)$ & $16.267(021)$ & \nodata \\ 
2010-05-07.71 & 55324.21 & $18.099(043)$ & \nodata & $16.394(027)$ & $16.369(030)$ \\ 
2010-05-11.74 & 55328.24 & $18.169(046)$ & \nodata & \nodata & \nodata \\ 
2010-05-18.60 & 55335.10 & $18.272(045)$ & \nodata & \nodata & \nodata \\ 
2010-06-04.62 & 55352.12 & $18.376(048)$ & \nodata & \nodata & \nodata \\ 
2010-06-06.60 & 55354.10 & \nodata & \nodata & $16.980(032)$ & $16.983(050)$ \\ 
2010-06-11.59 & 55359.09 & \nodata & $17.496(033)$ & $17.177(033)$ & $17.300(061)$ \\ 
2010-06-22.52 & 55370.02 & \nodata & $17.644(040)$ & $17.383(043)$ & \nodata \\ 
\enddata 
\tablecomments{Uncertainties in parentheses are given in thousandth of a magnitude.}
\end{deluxetable} 
 % Table 4

\clearpage
\begin{deluxetable}{lccccc} 
%\tabletypesize{\scriptsize} 
\tablecolumns{6} 
\tablewidth{0pt} 
\tablecaption{2MASS-system $JHK$ photometric sequence of field stars around SN~2010as.\label{tab:nseq}} 
\tablehead{ 
\colhead{Star} & \colhead{RA} & \colhead{Dec.} & \colhead{$J$} & \colhead{$H$} & \colhead{$K$} \\ 
\colhead{ID} & \colhead{(h:m:s)} & \colhead{($^\circ$:$'$:$''$)} & \colhead{(mag)} & \colhead{(mag)} & \colhead{(mag)} 
} 
\startdata
2  & 15:49:50.596 & $-29$:21:30.22 & $14.588(024)$ & $14.238(044)$ & $14.075(058)$ \\ 
8  & 15:49:44.342 & $-29$:23:52.40 & $14.859(047)$ & $14.485(061)$ & $14.471(083)$ \\ 
10 & 15:49:52.330 & $-29$:20:26.20 & $14.694(041)$ & $14.298(049)$ & $14.118(064)$ \\ 
11 & 15:49:53.861 & $-29$:20:42.09 & $15.619(066)$ & $14.850(067)$ & $14.788(106)$ \\ 
14 & 15:49:47.942 & $-29$:24:53.56 & $15.139(053)$ & $14.604(055)$ & $14.399(081)$ \\ 
15 & 15:49:46.103 & $-29$:25:10.48 & $15.472(073)$ & $15.228(109)$ & $14.918(134)$ \\ 
16 & 15:49:39.244 & $-29$:24:13.54 & $14.931(040)$ & $14.466(058)$ & $14.356(078)$ \\ 
18 & 15:49:39.608 & $-29$:20:27.26 & $14.440(038)$ & $14.117(048)$ & $14.019(065)$ \\ 
19 & 15:49:37.335 & $-29$:20:44.98 & $15.366(051)$ & $14.824(068)$ & $14.815(118)$ \\ 
\enddata 
\tablecomments{Uncertainties in parentheses are given in thousandth of a magnitude.}
\end{deluxetable} 
 % Table 5

\clearpage
\begin{deluxetable}{lcccc} 
\tabletypesize{\small} 
\tablecolumns{5} 
\tablewidth{0pt} 
\tablecaption{2MASS-system $JHK$ magnitudes of SN~2010as from GROND.\label{tab:nsn}} 
\tablehead{ 
\colhead{UT date} & \colhead{JD} & \colhead{$J$} & \colhead{$H$} & \colhead{$K$} \\ 
\colhead{(yyyy-mm-dd)} & \colhead{($-2400000$)} & \colhead{(mag)} & \colhead{(mag)} & \colhead{(mag)} 
} 
\startdata
2010-03-21.89 & 55277.39 & $15.385(055)$ & $15.607(068)$ & $15.774(152)$ \\ 
2010-03-22.85 & 55278.35 & $15.149(025)$ & $15.130(046)$ & $15.302(082)$ \\ 
2010-03-24.89 & 55280.39 & $15.097(017)$ & $14.982(030)$ & $15.082(071)$ \\ 
2010-03-26.91 & 55282.41 & $14.956(016)$ & $14.798(023)$ & $14.804(056)$ \\ 
2010-03-29.91 & 55285.41 & $14.667(012)$ & $14.637(020)$ & $14.513(044)$ \\ 
2010-04-01.85 & 55288.36 & $14.512(014)$ & $14.489(018)$ & $14.413(043)$ \\ 
2010-04-04.82 & 55291.32 & \nodata & \nodata & $14.296(042)$ \\ 
2010-04-06.92 & 55293.42 & $14.413(013)$ & $14.253(020)$ & $14.352(043)$ \\ 
2010-04-09.92 & 55296.42 & $14.384(013)$ & $14.245(015)$ & $14.286(037)$ \\ 
2010-04-10.91 & 55297.41 & $14.374(011)$ & $14.235(016)$ & $14.374(039)$ \\ 
2010-04-13.70 & 55300.20 & $14.468(013)$ & $14.246(017)$ & \nodata \\ 
2010-04-16.72 & 55303.22 & $14.680(061)$ & $14.399(094)$ & \nodata \\ 
2010-04-17.74 & 55304.24 & $14.623(014)$ & $14.333(024)$ & \nodata \\ 
2010-04-20.90 & 55307.40 & \nodata & $14.512(019)$ & \nodata \\ 
2010-04-23.92 & 55310.42 & $14.812(020)$ & $14.499(023)$ & $14.422(052)$ \\ 
2010-04-25.91 & 55312.42 & \nodata & $14.605(021)$ & $14.441(039)$ \\ 
2010-04-28.92 & 55315.42 & \nodata & $14.659(023)$ & $14.588(051)$ \\ 
2010-05-02.92 & 55319.42 & \nodata & \nodata & $14.781(057)$ \\ 
2010-05-07.93 & 55324.43 & $15.270(024)$ & $14.967(038)$ & $14.800(050)$ \\ 
2010-05-10.87 & 55327.37 & $15.456(023)$ & $15.011(030)$ & $15.112(060)$ \\ 
2010-05-19.76 & 55336.26 & $15.645(033)$ & $15.266(036)$ & $15.634(126)$ \\ 
2010-05-22.71 & 55339.21 & $15.650(038)$ & $15.357(049)$ & $15.585(117)$ \\ 
2010-05-29.70 & 55346.19 & $15.980(045)$ & $15.674(065)$ & $15.685(110)$ \\ 
2010-06-01.71 & 55349.21 & $16.003(060)$ & $15.757(087)$ & $15.735(164)$ \\ 
2010-06-04.55 & 55352.06 & $16.040(059)$ & $16.001(097)$ & $15.857(144)$ \\ 
2010-06-11.70 & 55359.19 & $16.210(058)$ & $16.106(100)$ & $16.174(289)$ \\ 
2010-06-13.46 & 55360.96 & $16.042(067)$ & $15.900(110)$ & \nodata \\ 
2010-06-21.46 & 55368.96 & $16.273(096)$ & $16.322(168)$ & \nodata \\ 
2010-06-29.58 & 55377.08 & $16.442(107)$ & $16.246(121)$ & $16.756(259)$ \\ 
2010-07-11.49 & 55388.99 & $16.610(072)$ & \nodata & \nodata \\ 
2010-07-19.46 & 55396.96 & $17.168(143)$ & $16.945(281)$ & \nodata \\ 
2010-07-29.47 & 55406.97 & $17.245(153)$ & \nodata & \nodata \\ 
2010-09-14.48 & 55453.98 & $17.240(217)$ & \nodata & \nodata \\ 
2011-02-17.89 & 55610.39 & $18.163(328)$ & \nodata & \nodata \\ 
\enddata 
\tablecomments{Uncertainties in parentheses are given in thousandth of a magnitude.}
\end{deluxetable} 
 % Table 6

\clearpage
\begin{deluxetable}{lcccccc}  % Table 7
\tabletypesize{\small} 
\tablecolumns{7} 
\tablewidth{0pt} 
\tablecaption{Spectroscopic observations of SN~2010as.\label{tab:speclog}} 
\tablehead{ 
\colhead{Date} & \colhead{JD} & \colhead{Epoch} & \colhead{Instrument} & \colhead{Wavelength} & \colhead{Airmass} & \colhead {Exposure} \\
\colhead{} & \colhead{$-2455000$} & \colhead{(days)} & \colhead{} & \colhead{Range (nm)} & \colhead{} & \colhead{(s)} \\
\colhead{(1)} & \colhead{(2)} & \colhead{(3)} & \colhead{(4)} & \colhead{(5)} & \colhead{(6)} & \colhead{(7)} 
}
\startdata 
2010-03-20 & $275.80$ & $-10.50$ & WFCCD     & $368-926$  & $1.062$ & 2400 \\
2010-03-21 & $276.90$ & $-9.41$  & GMOS-S    & $520-949$  & $1.028$ & 600  \\
2010-03-22 & $277.85$ & $-8.47$  & WFCCD     & $360-949$  & $1.002$ & 1800 \\
2010-03-23 & $278.82$ & $-7.51$  & WFCCD     & $360-949$  & $1.025$ & 1800 \\
2010-03-24 & $279.83$ & $-6.50$  & WFCCD     & $360-949$  & $1.011$ & 1800 \\
2010-03-25 & $280.83$ & $-5.50$  & X-Shooter & $320-2500$ & $1.005$ & 600  \\
2010-03-25 & $280.84$ & $-5.50$  & WFCCD     & $368-926$  & $1.003$ & 2700 \\
2010-03-26 & $281.85$ & $-4.49$  & X-Shooter & $320-2500$ & $1.008$ & 480  \\
2010-03-28 & $283.83$ & $-2.53$  & X-Shooter & $320-2500$ & $1.005$ & 480  \\
2010-03-29 & $284.73$ & $-1.64$  & X-Shooter & $320-2500$ & $1.215$ & 600  \\
2010-04-09 & $295.67$ & $9.22$   & GHTS      & $358-889$  & $1.37$  & 600  \\
2010-04-22 & $308.16$ & $21.62$  & GMOS-S    & $358-963$  & $1.20$  & 900  \\
2010-07-08 & $385.71$ & $98.61$  & GMOS-S    & $358-963$  & $1.50$  & 2000 \\
2010-07-23 & $400.60$ & $113.39$ & GHTS      & $353-885$  & $1.07$  & 1800 \\
2010-08-02 & $410.55$ & $123.27$ & GMOS-S    & $358-895$  & $1.06$  & 1800 \\
2010-08-05 & $413.48$ & $126.17$ & GHTS      & $354-886$  & $1.0$   & 900  \\
2011-02-05 & $597.87$ & $309.23$ & IMACS     & $398-1006$ & $1.244$ & 1800 \\
\enddata 
\tablecomments{Columns: (1) UT date of observation; (2) Julian date of
  middle of observation; (3) Epoch in rest-frame days since the time of
  $B$-band maximum light; (4) Instrument name; (5) Wavelength range of
  spectrum in nm; (6) Airmass at the mid-point of the observation; (7)
  Total exposure time in seconds.
}
\end{deluxetable} 

\clearpage
\begin{deluxetable}{lccccc}  % Table 8
\tabletypesize{\small} 
\tablecolumns{6} 
\tablewidth{0pt} 
\tablecaption{Light-curve parameters of SN~2010as derived from polynomial fits.\label{tab:lcfits}} 
\tablehead{ 
\colhead{Filter} & \colhead{JD$_{\mathrm{max}}$} & \colhead{$m_{\mathrm{max}}$} & \colhead{rms} & \colhead{Npt} & \colhead{$M_{\mathrm{max}}$} \\
\colhead{} & \colhead{$-2455000$} & \colhead{(mag)} & \colhead{(mag)} & \colhead{} & \colhead{(mag)} \\
\colhead{(1)} & \colhead{(2)} & \colhead{(3)} & \colhead{(4)} & \colhead{(5)} & \colhead{(6)} 
}
\startdata 
$B$  & $286.38 \pm 0.5$ &  $16.583 \pm  0.018$ & $0.025$ & 20 & $-17.26 \pm 0.44$ \\
$V$  & $288.78 \pm 0.5$ &  $15.667 \pm  0.009$ & $0.019$ & 19 & $-17.59 \pm 0.39$ \\
$R$  & $290.89 \pm 0.5$ &  $15.157 \pm  0.008$ & $0.026$ & 19 & $-17.85 \pm 0.38$ \\
$I$  & $292.40 \pm 0.5$ &  $14.837 \pm  0.008$ & $0.022$ & 18 & $-17.85 \pm 0.37$ \\
$g'$ & $287.37 \pm 0.5$ &  $16.138 \pm  0.010$ & $0.012$ & 19 & $-17.49 \pm 0.42$ \\
$r'$ & $289.87 \pm 0.5$ &  $15.397 \pm  0.006$ & $0.014$ & 21 & $-17.68 \pm 0.38$ \\
$i'$ & $291.69 \pm 0.5$ &  $15.298 \pm  0.007$ & $0.025$ & 21 & $-17.50 \pm 0.37$ \\
$z'$ & $291.09 \pm 0.5$ &  $15.324 \pm  0.010$ & $0.021$ & 20 & $-17.23 \pm 0.36$ \\
$J$  & $295.03 \pm 0.5$ &  $14.400 \pm  0.010$ & $0.064$ & 15 & $-17.99 \pm 0.36$ \\
$H$  & $297.74 \pm 1.0$ &  $14.250 \pm  0.011$ & $0.039$ & 17 & $-18.06 \pm 0.36$ \\
$K$  & $298.75 \pm 1.0$ &  $14.256 \pm  0.034$ & $0.091$ & 20 & $-18.00 \pm 0.36$ \\
 & & & & & \\
%\hline
Bol. & $287.99 \pm 0.5$ &  $14.678 \pm  0.005$ & $0.013$ & 32 & $-17.48 \pm 0.36$ \\
\enddata 
\tablecomments{Columns: (1) Filter name; (2) Julian date of maximum
  light. Uncertainties were estimated from the data sampling; (3)
  Apparent magnitude at maximum light, with no extinction correction
  applied. Uncertainties were computed from uncertainties in the fit
  parameters; (4) Root-mean-square of the data points around the fit;
  (5) Number of data points used in the fit; (6) Absolute peak
  magnitude in each band $X$,
  $M_{\mathrm{max}}(X)=m_{\mathrm{max}}(X)\,-\,A_{\mathrm{Gal}}(X)\,-\,A_{\mathrm{host}}(X)\,-\,\mu$,
  with Galactic extinction $A_{\mathrm{Gal}}(X)$ from
  $E(B-V)_{\mathrm{Gal}}=0.151$ mag and $R^{\mathrm{Gal}}_V=3.1$,
  host-galaxy extinction $A_{\mathrm{host}}(X)$ from
  $E(B-V)_{\mathrm{host}}=0.42\pm0.1$ mag and $R^{\mathrm{host}}_V=1.5$,
  and distance modulus $\mu=32.16\pm0.36$ mag. 
}
\end{deluxetable} 

\clearpage
\begin{deluxetable}{lccccc} 
\tabletypesize{\scriptsize} 
\tablecolumns{6} 
\tablewidth{0pt} 
\tablecaption{Black-body fits and bolometric light curve of SN~2010as.\label{tab:lbol}} 
\tablehead{ 
\colhead{UT date} & \colhead{JD} & \colhead{$T_{\mathrm{BB}}$} & \colhead{$R_{\mathrm{BB}}$} & \colhead{$\log(L_{B\rightarrow K})$\tablenotemark{a}} & \colhead{$\log(L_{\mathrm{Bol}})$\tablenotemark{a}} \\ 
\colhead{(yyyy-mm-dd)} & \colhead{($-2400000$)} & \colhead{(K)} & \colhead{($10^{15}$ cm)} & \colhead{(erg s$^{-1}$)} & \colhead{(erg s$^{-1}$)} 
} 
\startdata
2010-03-19.70 & $55275.20$ & $10023 \pm   895$ & $ 0.70 \pm  0.08$ & $41.75$ &  $41.87$ \\ 
2010-03-20.70 & $55276.20$ & $12013 \pm   919$ & $ 0.62 \pm  0.05$ & $41.83$ &  $41.97$ \\ 
2010-03-21.70 & $55277.20$ & $10063 \pm   604$ & $ 0.86 \pm  0.06$ & $41.92$ &  $42.06$ \\ 
2010-03-22.85 & $55278.35$ & $ 8659 \pm   409$ & $ 1.15 \pm  0.06$ & $41.99$ &  $42.14$ \\ 
2010-03-23.70 & $55279.20$ & $ 8923 \pm   429$ & $ 1.17 \pm  0.06$ & $42.04$ &  $42.18$ \\ 
2010-03-24.70 & $55280.20$ & $ 9296 \pm   461$ & $ 1.18 \pm  0.06$ & $42.09$ &  $42.24$ \\ 
2010-03-25.70 & $55281.20$ & $ 9423 \pm   472$ & $ 1.22 \pm  0.06$ & $42.14$ &  $42.29$ \\ 
2010-03-26.73 & $55282.22$ & $ 9207 \pm   445$ & $ 1.30 \pm  0.06$ & $42.16$ &  $42.32$ \\ 
2010-03-27.70 & $55283.20$ & $ 9171 \pm   437$ & $ 1.35 \pm  0.06$ & $42.19$ &  $42.35$ \\ 
2010-03-28.74 & $55284.24$ & $ 9148 \pm   431$ & $ 1.41 \pm  0.07$ & $42.22$ &  $42.38$ \\ 
2010-03-29.70 & $55285.20$ & $ 8938 \pm   405$ & $ 1.49 \pm  0.07$ & $42.24$ &  $42.40$ \\ 
2010-03-30.70 & $55286.20$ & $ 8984 \pm   405$ & $ 1.52 \pm  0.07$ & $42.26$ &  $42.41$ \\ 
2010-03-31.70 & $55287.20$ & $ 8827 \pm   388$ & $ 1.57 \pm  0.07$ & $42.28$ &  $42.42$ \\ 
2010-04-01.72 & $55288.22$ & $ 8718 \pm   373$ & $ 1.62 \pm  0.07$ & $42.29$ &  $42.43$ \\ 
2010-04-02.70 & $55289.20$ & $ 8421 \pm   347$ & $ 1.69 \pm  0.07$ & $42.29$ &  $42.43$ \\ 
2010-04-03.70 & $55290.20$ & $ 8038 \pm   306$ & $ 1.80 \pm  0.07$ & $42.29$ &  $42.41$ \\ 
2010-04-04.71 & $55291.21$ & $ 7676 \pm   276$ & $ 1.90 \pm  0.08$ & $42.28$ &  $42.40$ \\ 
2010-04-05.70 & $55292.20$ & $ 7460 \pm   255$ & $ 1.96 \pm  0.08$ & $42.27$ &  $42.38$ \\ 
2010-04-06.71 & $55293.21$ & $ 7248 \pm   236$ & $ 2.02 \pm  0.08$ & $42.26$ &  $42.37$ \\ 
2010-04-07.70 & $55294.20$ & $ 6992 \pm   216$ & $ 2.10 \pm  0.08$ & $42.25$ &  $42.35$ \\ 
2010-04-08.71 & $55295.21$ & $ 6744 \pm   198$ & $ 2.18 \pm  0.08$ & $42.23$ &  $42.33$ \\ 
2010-04-09.70 & $55296.20$ & $ 6503 \pm   182$ & $ 2.26 \pm  0.08$ & $42.22$ &  $42.31$ \\ 
2010-04-10.70 & $55297.20$ & $ 6440 \pm   176$ & $ 2.25 \pm  0.08$ & $42.20$ &  $42.29$ \\ 
2010-04-11.70 & $55298.20$ & $ 6392 \pm   175$ & $ 2.22 \pm  0.08$ & $42.17$ &  $42.26$ \\ 
2010-04-13.70 & $55300.20$ & $ 6315 \pm   169$ & $ 2.14 \pm  0.08$ & $42.13$ &  $42.21$ \\ 
2010-04-14.70 & $55301.20$ & $ 6414 \pm   184$ & $ 1.98 \pm  0.08$ & $42.10$ &  $42.19$ \\ 
2010-04-16.72 & $55303.22$ & $ 6360 \pm   189$ & $ 1.88 \pm  0.08$ & $42.06$ &  $42.14$ \\ 
2010-04-17.74 & $55304.24$ & $ 5997 \pm   150$ & $ 2.11 \pm  0.07$ & $42.05$ &  $42.13$ \\ 
2010-04-18.70 & $55305.20$ & $ 5849 \pm   143$ & $ 2.16 \pm  0.08$ & $42.04$ &  $42.12$ \\ 
2010-04-20.70 & $55307.20$ & $ 5527 \pm   129$ & $ 2.33 \pm  0.08$ & $42.00$ &  $42.08$ \\ 
2010-04-23.92 & $55310.42$ & $ 5181 \pm   119$ & $ 2.50 \pm  0.09$ & $41.96$ &  $42.04$ \\ 
2010-04-25.70 & $55312.20$ & $ 5128 \pm   116$ & $ 2.47 \pm  0.08$ & $41.94$ &  $42.01$ \\ 
2010-05-07.93 & $55324.43$ & $ 5230 \pm   127$ & $ 2.02 \pm  0.07$ & $41.78$ &  $41.87$ \\ 
2010-05-10.87 & $55327.37$ & $ 5472 \pm   135$ & $ 1.78 \pm  0.06$ & $41.74$ &  $41.83$ \\ 
2010-05-11.70 & $55328.20$ & $ 5442 \pm   143$ & $ 1.78 \pm  0.07$ & $41.73$ &  $41.82$ \\ 
2010-05-17.70 & $55334.20$ & $ 5700 \pm   154$ & $ 1.55 \pm  0.06$ & $41.68$ &  $41.77$ \\ 
2010-05-19.76 & $55336.26$ & $ 5722 \pm   161$ & $ 1.51 \pm  0.06$ & $41.66$ &  $41.75$ \\ 
2010-05-30.10 & $55346.60$ & $ 6052 \pm   197$ & $ 1.22 \pm  0.06$ & $41.56$ &  $41.65$ \\ 
2010-06-05.10 & $55352.60$ & $ 6349 \pm   238$ & $ 1.06 \pm  0.06$ & $41.51$ &  $41.61$ \\ 
2010-06-12.00 & $55359.50$ & $ 6440 \pm   265$ & $ 0.97 \pm  0.06$ & $41.45$ &  $41.55$ \\ 
2010-06-22.52 & $55370.02$ & $ 6973 \pm   344$ & $ 0.75 \pm  0.06$ & $41.38$ &  $41.47$ \\ 
2010-07-12.00 & $55389.50$ & $ 6275 \pm   508$ & $ 0.82 \pm  0.10$ & $41.23$ &  $41.32$ \\ 
\enddata 
\tablenotetext{a}{A distance-dominated systematic uncertainty of
  $\approx$36\% in luminosity can be considered.}
\end{deluxetable} 
 % Table 9

\clearpage
\begin{deluxetable}{lcccll}  % Table 10
\tabletypesize{\small} 
\tablecolumns{6} 
\tablewidth{0pt} 
\tablecaption{Properties of the flat-velocity Type~IIb SNe.\label{tab:flatIIb}} 
\tablehead{ 
\colhead{SN} & \colhead{Orig.} & \colhead{$v_{\mathrm{He}}({\mathrm{max}})$} & \colhead{$M_V({\mathrm{max}})$} & \colhead{Type} & \colhead{Phot.} \\
\colhead{name} & \colhead{type} & \colhead{($10^3$ km s$^{-1}$)} & \colhead{(mag)} & \colhead{refs.} & \colhead{refs.} \\
\colhead{(1)} & \colhead{(2)} & \colhead{(3)} & \colhead{(4)} & \colhead{(5)} & \colhead{(6)} 
}
\startdata 
1999ex & Ib/c      & $6.93 \pm 0.21$ & $-17.49 \pm 0.40$ & $a$ & $b$ \\
2005bf & Ib/c      & $6.31 \pm 0.08$ & $-17.40 \pm 0.31$ & $c$ & $c$ \\
2007Y  & Ib/c, IIb & $8.23 \pm 0.38$ & $-16.38 \pm 0.34$ & $d$, $e$ & $d$\\
2010as & \nodata   & $5.41 \pm 0.12$ & $-17.59 \pm 0.39$ & \nodata & \nodata \\
2011dh & IIb       & $7.24 \pm 0.14$ & $-17.08 \pm 0.34$ & $f$, $g$ & $g$ \\
\enddata 
\tablecomments{Columns: (1) SN name; (2) Original classification from
  the literature (when different types were found, they are all listed); (3)
  Velocity of the \ion{He}{1}~$\lambda$5876 line at $B$-band maximum
  light. Values obtained from straight-line fits to the data in the
  epoch range of $[-10:+10]$ days; (4) Reddening-corrected absolute peak
  magnitude in $V$ band. Distances and Galactic reddening adopted from
  NED, apparent peak magnitudes and host-galaxy reddening adopted from
  the references given in column (6); (5) References for the spectral
  typing of column (2); (6) References for the photometric information
  used to calculate $M_V({\mathrm{max}})$. References: $a$ -
  \citet{hamuy02}; $b$ - \citet{stritzinger02}; $c$ -
  \citet{folatelli06}; $d$ - \citet{stritzinger09}; $e$ -
  \citet{maurer10}; $f$ - \citet{arcavi11}; $g$ - \citet{ergon14}.
}
\end{deluxetable} 

\end{document}